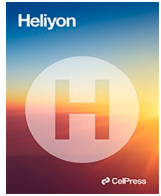

Review article

# 6G communications through sub-Terahertz CMOS power amplifiers: Design challenges and trends

Jun Yan Lee [a,b], Duo Wu [a], Xuanrui Guo [a], Jian Ding Tan [a,*], Teh Jia Yew [c,**],
Zi Neng Ng [a], Mohammad Arif Sobhan Bhuiyan [a], Mahdi H. Miraz [c,d,e,***]

[a] *School of Artificial Intelligence and Robotics, Xiamen University Malaysia, Selangor, Malaysia*
[b] *School of Electronic, Information and Electrical Engineering, Shanghai Jiao Tong University, Shanghai, China*
[c] *School of Computing and Data Science, Xiamen University Malaysia, Selangor, Malaysia*
[d] *School of Computing, Faculty of Arts, Science and Technology, Wrexham University, Wrexham, UK*
[e] *Faculty of Computing, Engineering and Science, University of South Wales, Pontypridd, UK*

**ABSTRACT**

The fifth-generation (5G) network faces limitations in supporting emerging applications, such as artificial intelligence (AI), virtual reality (VR) and digital twins. To overcome these confines, sub-Terahertz (sub-THz) and Terahertz (THz) technologies are considered to be key enablers of effective 6G wireless communications, offering higher transmission speeds, longer range and wider bandwidth. Achieving these capabilities requires careful engineering of 6G transceivers, with a focus on efficient power amplifiers (PAs) in the front-end, which play a critical role in effectively amplifying and transmitting signals over long distances. Complimentary metal-oxide-semiconductor (CMOS) technology-based PA in sub-THz suffers severe parasitic and limited maximum frequency, however, this has eventually been solved by different design architectures and scaling down of CMOS technology to break through the frequency limitations. In this article, we reviewed the potentials and capabilities of CMOS technology for designing 6G hardware, identified the state-of-art PA designs in the sub-THz band and then examined as well as compared the designs to identify the suitable design strategies for better performance. The circuit optimisation techniques, such as coupled-line, passive gain boosting method, zero-degree power splitting, load-pull matching, diode and capacitor linearisation for better gain, saturated output power and power added efficiency, are considered for the PA design architectures at different sub-THz bands. Furthermore, these methods are summarised and discussed with their advantages and disadvantages in lieu with their performances. The PA design trends, challenges and future perspectives are also presented and discussed. Therefore, this comprehensive review article will serve as a comparative study and reference for future PA designs for radio frequency integrated circuits (RFIC).

---

* Corresponding author.
** Corresponding author.
*** Corresponding author. School of Computing and Data Science, Xiamen University Malaysia, Selangor, Malaysia.
   *E-mail addresses:* jianding.tan@xmu.edu.my (J.D. Tan), jiayew.teh@xmu.edu.my (T.J. Yew), m.miraz@ieee.org (M.H. Miraz).

## 1. Introduction

With the evolution of the wireless communications systems since 1980s, the fifth generation (5G) communications system has recently been released and it has been dominating in terms of low latency, high transmission rate as well as enhanced quality of service (QoS) [1]. That being said, the 5G communications system still has several constraints, particularly, in terms of bandwidth and transmission rate, specifically when it comes to the next generation virtual communications applications [2]. Therefore, advancing towards the sixth generation (6G) communications system becomes essential to further enhance the service and capabilities of the communication networks. The advent of 6G holds promises to unlock various cutting-edge services, such as, augmented reality (AR)/virtual reality (VR) applications, digital twins, unmanned aerial vehicles, autonomous driving, etc. which demand supplemental transmission rate, latency and reliability than what the current 5G network can provide [3–9]. The sixth generation (6G) communications system is mapped out to be commercially released to the public by 2030. Additionally, 6G transmission bands covering sub-Terahertz (sub-THz) and the Terahertz (THz) frequencies with high data transfer rate ranging from 100 Giga bits per second (Gbps) to 1 Tera bits per second (Tbps), which is more than 10 times higher than that of the 5G communications system, is the key takeaway and area to actualise in 6G communications system. Table 1 points out the key performance indicators (KPIs) as well as the system requirements for the 6G communication network [10,11]. In 2017, the first standardisation effort i.e., IEEE 802.15.3d, was released. IEEE 802.15.3d outlines a high data transfer rate wireless network for fixed point-to-point application in sub-THz communication for 6G [12,13]. Two years later, in March 2019, the Federal Communications Commission (FCC) opens out the THz spectrum band ranging from 95 Gigahertz (GHz) to 3 THz for researchers and engineers to conduct experiments and unlicensed activities to facilitate developing 6G systems [14]. Several organisations including International Telecommunication Union [15], Keysight [16], Rohde and Schwarz [11], Samsung [17,18], Huawei [19] and Nokia [20] have published white papers outlining their visions, key research areas and proposed network requirements and KPIs of the new applications in THz band to provide high-speed communications with low throughput. China successfully launched its first 6G satellite in November 2020, which employed THz waves to try-out 6G technologies [21]. In March 2022, Keysight secured the first Federal Communications Commission (FCC) Spectrum Horizons License, particularly for the development of 6G technologies in Sub-THz frequency range [22]. These significant milestones mark the starts of a new era towards advanced sub-THz to THz band test-bed enabling innovative and cutting-edge research in 6G supported physical layer hardware.

To enable a 6G network communications system, the transceiver plays a very important role. While 6G is designed to operate at sub-THz and THz bands, these frequency bands have already been being widely developed since the 1980s, especially for THz imaging and sensing for the purpose of security, defence and medical applications [2,23]. Refer to block diagram presented in Fig. 1, a transceiver is made up of a transmitter (Tx), a receiver (Rx), as well as a T/R switch to allow the signal to perform frequency transformation, remove unwanted signals and adjust frequency levels while transmitting or receiving. Research in the sub-THz region marks a significant starting point before venturing in to the higher THz frequencies. A sub-THz power amplifier (PA) is employed at the transmitter's end. This core module plays a very important role in enabling signal transmissions by performing signal amplifications through several crucial factors, such as its gain, power consumption based on the efficiency as well as transmission distance, quantity quality based on the output power, bandwidth and linearity, respectively. In some cases, certain PA system characteristics take precedence over others depending on the applications. For instance, high-efficiency PA is pivotal in reducing power dissipation in the form of heat and addressing cooling requirements in base stations; the requirement for adequate output power supersedes that of the efficiency, particularly for long-range communication in noisy environments. However, PA demands substantial power and supply voltage for executing these functions. It is crucial to assure high linearity in order to minimise heat dissipation while maintaining high efficiency. Furthermore, designing PAs, operating at high frequencies, is challenged by existing technological limitations. Therefore, to address these complexities, the PA designs have adopted multifaceted technological processes, such as Indium Phosphide Heterojunction Bipolar Transistor (InP HBT) [24], Indium Phosphide High Electron Mobility Transistors (InP HEMT) [25], Gallium Nitride (GaN) [26], Gallium Arsenide metamorphic high electron mobility transistor (GaAs mHEMT) [27], Gallium Arsenide Pseudomorphic High Electron Mobility Transistors (GaAs pHEMT) [28], Complementary Metal-Oxide-Semiconductor (CMOS) [29] and Silicon Germanium Bipolar Complementary Metal-Oxide-Semiconductor (SiGe BiCMOS) [30], to evaluate the design performance and investigate the optimisation methods. A survey of the saturated output power from PA Monolithic Microwave Integrated Circuit (MMIC), recently demonstrated in different semiconductor technologies, is shown in Fig. 2. Amongst these technologies, CMOS stands out as it offers various advantages, such as cost-effectiveness, technology scaling and reduced power dissipation, while designing PAs

**Table 1**
Key performance indicators (KPI) enhancements from 5G to 6G [11].

| Key performance indicators (KPI) | 5G | 6G | Improvement factor |
|---|---|---|---|
| Peak data rate (Gbps) | 10 | 100 to 1000 | 10 to 100 |
| User-experienced data rate (Gbps) | 0.1 | 1 to 10 | 10 to 100 |
| User plane latency (ms) | 1 | 0.1 | 10 |
| Connection density (device/km$^2$) | $10^6$ | $10^7$ to $10^8$ | 10 to 100 |
| Reliability | 99.999 % | 99.99999 % | 100 % |
| Energy efficiency | $1 \times$ | $5 \times 100 \times$ | 5 to 100 |
| Spectral efficiency | $1 \times$ | $2 \times$ | 2 |
| Positioning (cm) | 20 to 100 in 2D | 1 in 3D | 20 to 100 |
| Jitter, i.e. latency variations (μs) | – | 0.1 to 1000 | – |

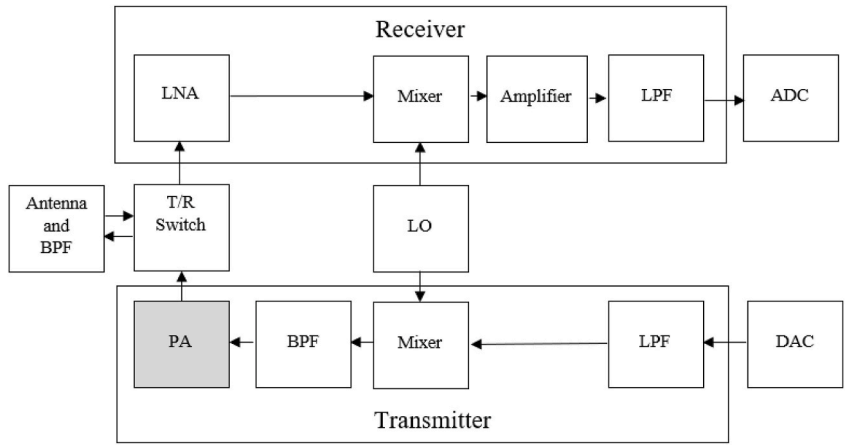

**Fig. 1.** Typical block diagram of a transceiver.

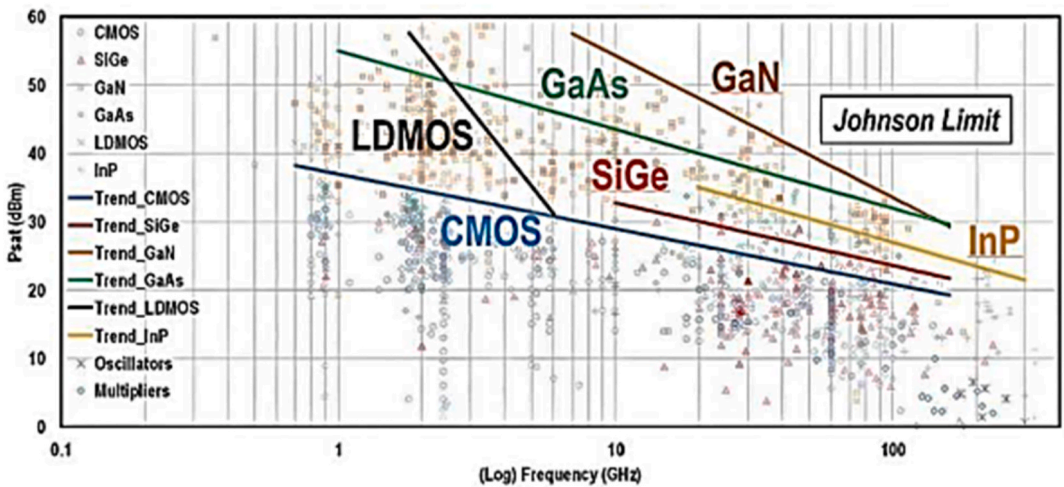

**Fig. 2.** PA saturated output power for different technologies versus frequencies [5].

and other radio frequency (RF) devices [31,32].

Nevertheless, CMOS technology has its own limitations such as high output conductance, transit frequency and velocity saturations. Such limitations of CMOS technology have comprehensively been presented in several research articles [32–35]. Scaling issue is currently the most serious concern when dealing with lower nodes CMOS technology, where shrinking transistors to sub-nanometre sizes face barriers, such as like leakage currents and quantum effects [32,33]. Increased power consumption and heat dissipation due to static and dynamic power losses also hinder further miniaturisation [34]. Economically, the high costs of manufacturing advanced nodes reduce the returns from following the Moore's Law [35]. Additionally, design complexity grows due to parasitic effects as well as interconnect delays; material limitations also make it harder to maintain performance at smaller scales [35]. Despite these inherent complexities, overcoming these issues from a design perspective is feasible but remain challenging. Therefore, designers must make careful informed choice about appropriate technology integration with various circuit design techniques to achieve their objectives. In fact, there are various circuit design techniques available in the sub-THz domain, such as: varieties of amplifiers, power combining and splitting, stages, configurations and neutralisation methods. As a matter of fact, PA stages are interconnected in different ways, such as: cascode, cascade, differential or pseudo-differential manner with PA common gate (CG), common gate common source (CG-CS), common source (CS) or common source-common source (CS-CS), each utilising district biasing as well as feedback techniques. These intricate technologies and design approaches aim to optimise PA performance in the sub-THz realm.

In this comprehensive literature review article, we delved into the potential PA designs for 6G communication technologies and highlighted various major facets of state-of-the-art PA designs in different CMOS processes, considering diverse performance parameters. Due to the scope limitations, the focus of this article primarily centres on sub-THz CMOS PAs, while highlighting both challenges and available circuit techniques. The remainder of this paper unfolds as follows: section 2 presents various CMOS technologies, highlighting their potentials and challenges while operating in the high-frequency sub-THz band; section 3 describes the system requirements, focusing on the essential PA parameters; section 4 reviews the landscape of available sub-THz CMOS PAs, including their design architectures, current state-of-the-arts achievements and notable advancements; Section 5 presents a comparative analysis of the PA architectures, discussing their performances, trends and ultimately identifying the optimal design based on each performance parameter; and lastly, section 6 concludes the article offering valuable insights into the present and future perspectives of this dynamic field of research and development.

Through this exploration, we aspire to contribute to the ongoing evolution of 6G communications system, paving the way for transformative advancements in wireless networks.

## 2. CMOS potentials and challenges

The development of semiconductor technology has played a very important role in shaping the communications systems over the years, from early days of 2G to the cutting-edge 6G. Radio Frequency (RF), CMOS capability and relative Power Amplifiers are mainly evaluated in terms of power generation, high frequency and high speed. In fact, when transitioning from 5G to 6G, a bottleneck occurs, since the speed of the transistor doesn't exhibit much change when moving from the lower Millimetre wave (mmW) region to a higher frequency part [16].

The performance of transistors, as active components as well as other passive components, is limited, when operating in upper mmW or sub-THz systems. However, certain types of devices, such as silicon and III-IV semiconductors have the ability to support applications for frequencies exceeding 100 GHz. The evolution of the CMOS processing node is illustrated in Fig. 3.

It has been observed that a transistor's intrinsic gain increases at a faster rate than its extrinsic gain [36]. In a digital circuit, the rising trend of the intrinsic gain implies that the closer the transistors are placed, the more benefit can be obtained from the intrinsic gain. However, extrinsic gain is important to ensure optimal performance in an RF analogue circuit. This is due to the resonator load, located on higher metal layers, used to terminate the amplifying transistor. One of the main challenges preventing RF devices from fully utilising the benefits of the latest semiconductor evolution is the interconnect issue. This further complicates the design process as the frequency increases. Two merit frequencies transit frequency ($f_T$) and maximum oscillation frequency ($f_{max}$) need to be evaluated. As a general guideline, the ratio of the two, i.e. $f_T/f_{max}$, need to be more than twice the operating frequency($f_O$), to achieve distinct performance of gain and efficiency in PA design. Although there are several existing technologies and applications, such as satellite communication, THz-sensors and THz biomedical radar operating in sub-THz and THz regions, it remains challenging to implement larger systems like RF transceivers compared to 5G frequencies in the lower mmW region. In fact, it still remains implausible to gain similar performance with the same level of power consumption even for 5G Frequency Range 1 (FR1) sub-6 GHz at lower mmW frequencies. This is due to the limitations of different semiconductor technologies from their physics and boundaries.

Current CMOS and CMOS SOI $f_T$ and $f_{max}$ saturate at 400 GHz, yet the $f_{max}$ of InP and GaAs transistors can exceed 1 THz [16]. Therefore, it is now feasible to realise mmW circuits in CMOS [37]. However, CMOS technology still lags in terms of power generation compared to silicon HBT transistors in SiGe BiCMOS and III-IV semiconductors. This is due to fact that CMOS process are optimized for digital circuits, which means passive elements, such as like inductors and capacitors in CMOS, have higher losses and poor performance at high frequencies. Fig. 4 illustrates a power amplifier with parasitic effects (e.g., parasitic capacitance and inductance) inherent in CMOS transistor. These parasitic effects form unintended low-pass filters, which limits the bandwidth, leads to signal attenuation, reduces gain and makes it difficult to achieve high-frequency operation. For instance, packaged 1-THz Tx/Rx chipsets with −37 dBm [38], single SiGe radiators deliver −6.3 dBm [39], or up to 10.3 dBm in an 8x8 array at 430 GHz [40]. The above discussion on InP, GaN, and CMOS devices demonstrates the capability of application in 6G communication, however, PAs' efficiencies are generally lower than 10 % [24,41–45]. Significant amount of research is being conducted, both in the academia and the industry, on different semiconductor technologies to better optimise their power generation capabilities in sub-THz and THz region.

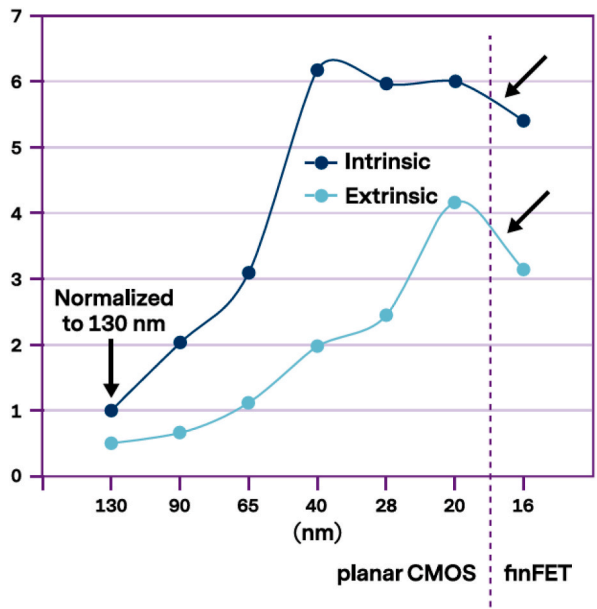

**Fig. 3.** Normalised intrinsic and extrinsic device FT [16].

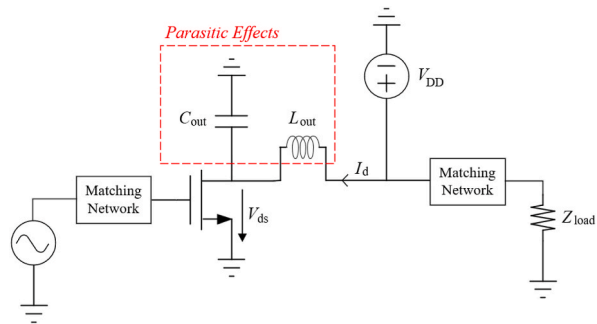

**Fig. 4.** Circuit diagram of a parasitic effect on the transistor of a basic power amplifier model.

## 3. Design parameters of PAs

The four most important parameters of a PA are: gain, linearity, saturated output power and power added efficiency (PAE). In PA design, the saturated output power determines the signals' transmission range. On the contrary, the amplification gain plays an important role in boosting the signals for demodulation, when received at the receiver side. The linearity of the signal also depends on the PA design to ensure that the amplified signal can be transmitted in a linear form. While the PAE is a key parameter in determining the power usage and the efficiency of the PA, this trade-off adds further complexity and challenges to the PA design process. The details of the parameters used to analyse the performance of a PA are provided in the following sub-sections.

### 3.1. Spectrum considerations

The range of the operating frequency of a PA can span from a several hundred MHz to THz accommodating different wavelengths, as shown in Fig. 5. In fact, the mmW and THz frequency bands are now the prime choice for developing 6G systems. FCC also unlocked the THz spectrum band between 95 GHz to 3 THz, particularly for research and develop of 6G systems [14]. The two components of the THz spectrum band include: the sub-THz range from 95 GHz to 300 GHz as well as the THz range from 300 GHz to 3 THz. It is anticipated that using a high frequency of THz would meet the needs of low latencies, high transmission rate and high data throughput speed from 100 Gbps to 1 Tbps. W- and D-band are also considered as potential candidates to be allocated for mobile services, as depicted in Fig. 4 [18]. While designing a 6G PA, potential operation at a minimum of sub-THz frequency should be considered, as it holds significant prospects to be deployed across numerous Institute of Electrical and Electronics Engineers (IEEE) frequency bands, as demonstrated in Table 2 [46].

### 3.2. Power gain

The ratio of output power to the input power is considered as the power gain, as articulated in eq. (1). Scattering parameter S21 also represents the power gain, as expressed in eq. (2). In fact, the power gain is a crucial parameter in measuring the increase of signal amplitude due to amplification. Generally, the gain reduces when the operating frequency increases, therefore, the contribution from the input power drive is important to be considered [47]. Especially, the PA design needs to maintain a good power gain in a high-frequency spectrum for 6G applications. A cross-connected neutralisation capacitor on a differential pair is also widely adopted to improve the stability of each PA stage and enhance the isolation of the modulated input as well as output signals.

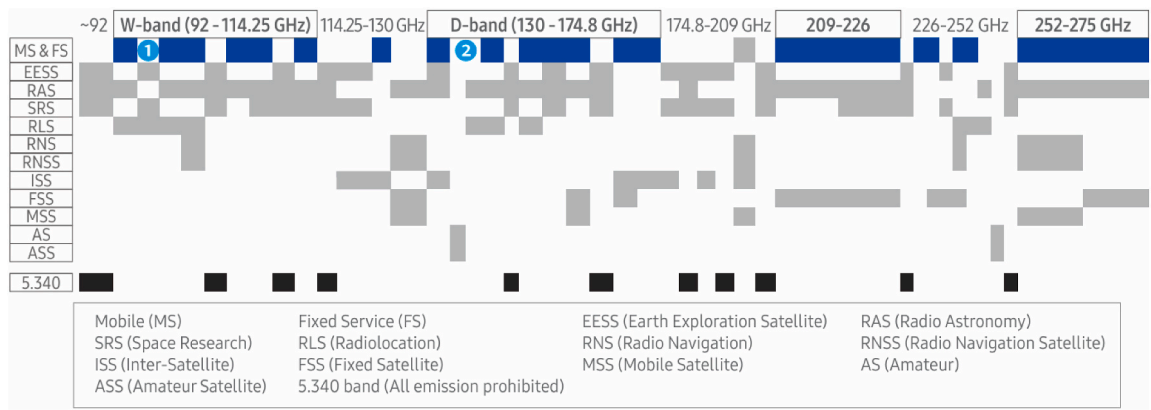

**Fig. 5.** Global allocation for 92–275 GHz spectrum range [18].

**Table 2**
IEEE frequency bands [46].

| Band | Frequency Range | Wavelength |
|------|-----------------|------------|
| HF | 3 MHz–30 MHz | 100 m - 10m |
| VHF | 30 MHz–300 MHz | 10 m–1 m |
| UHF | 300 MHz–1000 MHz | 1 m - 10 cm |
| L | 1 GHz–2 GHz | 30 cm–15 cm |
| S | 2 GHz–4 GHz | 15 cm–7.5 cm |
| C | 4 GHz–8 GHz | 7.5 cm–3.75 cm |
| X | 8 GHz–12 GHz | 3.75 cm–2.5 cm |
| Ku | 12 GHz–18 GHz | 2.5 cm–1.6 cm |
| K | 18 GHz–27 GHz | 1.6 cm–1.2 cm |
| Ka | 27 GHz–40 GHz | 1.2 cm - 7.5 mm |
| V | 40 GHz–75 GHz | 7.5 mm–4 mm |
| **W** | **75 GHz - 110 GHz** | 4 mm - 2.8 mm |
| **D** | **110 GHz - 170 GHz** | 2.8 mm – 1.8 mm |
| **mm** | **110 GHz - 300 GHz** | 2.8 mm – 1 mm |

$$\text{Gain} = 10 \; log_{10}\left(P_{out,RF} \; / \; P_{in,RF}\right)[dB] \tag{1}$$

$$\text{Gain} = 20 \; log_{10}|S21|[dB] \tag{2}$$

### 3.3. Output power

Output power is the quantity of power delivered to the antenna for transmission, expressed by eq. (3). It is an important parameter determining the transmission distance, as higher output power is able to transmit to a longer distance. In fact, this parameter is closely related to the PA Design's power gain and efficiency (PAE & DE), which significantly impact the performance of the PA.

$$P_{out} = \frac{V_{out}}{2R_L} \tag{3}$$

The effective isotropic radiated power (EIRP) is taken into account to estimate the radiated output power of an antenna. The EIRP requirements vary across applications, such as the handset and the base station. Examples of EIRP requirements for different links, the number of antennas needed as well as the PA requirements – all are illustrated in Table 3 [16]. For a base station, the antenna array consists of 256 elements. A 75 dBm ERIP and an average output power of 25–27 dBm per PA is required. The design of a handset is complicated and is dependent on the number of antennas [48]. Considering the number of arrays of antenna consisting of 16 elements, a 45 dBm ERIP and the average output power of 16–20 dBm per PA is required.

### 3.4. Efficiency

The ratio of RF output power to the DC power, as expressed in percentage in eq. (4), is known as drain efficiency (DE), as the conversion of DC power to RF power occurs at the output. Another efficiency parameter of the power-added efficiency (PAE) is the ratio of RF output power and the gate-driven power to the DC supply power, as expressed in percentage in eq. (5). The PAE of the PA reaches its highest when the peak output power (PEP) is achieved and it drops with decline of the output power. Similarly, the efficiency of PA can be maximised by ensuring that the power dissipation at the output power is minimised [49]. When the transistor of the PA operates as a current source, DE is observed at the peak envelope power. Furthermore, a power backoff will also occur, as the PA will operate below the maximum output power [49]. A high peak to average power ratio (PAPR) signal, i.e. the ratio of the peak power to the average power of the signal, implies a short operating time of the PA at its point of peak efficiency. It is very important to reduce the PAPR, in order to allow the PA to operate a longer period of time at its peak efficiency.

**Table 3**
Application Scenarios and estimated requirements for 6G > 100 GHz Transmitter [16].

|  | Handset | Base station |
|--|---------|--------------|
| EIRP | 45 dBm | 75 dBm |
| $N_{ant}$ | 16 | 256 |
| $P_{ave}$/PA | 16 dBm – 20 dBm | 25 dBm – 27 dBm |

$$DE = 100 \times \left(P_{out,RF} / P_{DC}\right) \tag{4}$$

$$PAE = 100 \times \left(P_{out,RF} / P_{in,RF}\right) / P_{DC} \tag{5}$$

### 3.5. Linearity

The need for linearity is important when designing a PA. When a signal, in a hybrid modulation, encompasses both amplitude modulation (AM) as well as phase modulation (PM), it is crucial to have linear amplification in the design. The nonlinearity of PA will cause spectral regrowth and degradation of spectral purity which will produce two unwanted signals, viz., harmonics distortion (HD) of carrier frequency, as expressed in eq. (6) and third-order intermodulation distortion (IDM), as expressed in eq. (7). A high linearity of the PA indicates that its output power is linear to the input power. There exist several methods to measure the nonlinearity of PA, such as adjacent channel power ratio (ACPR), error vector magnitude (EVM), carrier-to-intermodulation (C/I) ratio and noise-power ratio (NPR). Amongst these methods, ACPR is widely used to measure linearity.

$$f_h = nf_c \tag{6}$$

$$f_{IDM} = nf_1 \pm mf_2 \tag{7}$$

## 4. Design topologies of PAs

Fig. 6 illustrates the basic block diagram of a CMOS power amplifier (PA). A basic CMOS PA possesses two main stages which are both biased separately, namely, a driver stage and a power stage. The input as well as the output matching networks are necessary for reducing the return loss, thus improving the gain and the output power. This also applies to the inter-stage matching between the driver and the power stages.

The PA consists of a variety of circuit architectures to suit the needs of various applications such as point-to-point, radar, satellite communications, and THz imaging. Power gain, power added efficiency (PAE), saturated output power and bandwidth are the important factors when designing a PA. The PA can be categorised into different architectures which operate in the sub-THz frequency band, i.e., the potential band for future 6G communications. Firstly, the PAs available with single- or multi-stages can be designed in differential or single-ended topology. Common-source (CS) maintains the most common topology while there are still few designs with common-gate (CG)-like topology, particularly when stack or cascode topology is used. The PA design also includes various techniques such as neutralisation capacitor, input-matching, lineariser, power splitting and combining. Fig. 7 illustrates the key topologies that successfully designed the sub-THz CMOS PA. Overall, the PA topologies need to be modified in order to introduce the best design outcomes addressing the requirements set by the application. Various PA architectures with their respective circuit diagrams and discussed in the following sub-sections.

### 4.1. IQ current-combining with differential class A PA

The PA circuit design, presented by Elazar et al. [50], as given in Fig. 8, offers a 2-way IQ current combined PA design to control 360° of the output phases. The core PA consists of three pseudo-differential CS stages, coupled by a stacked inductors transformer, for better SRF and cross-coupled neutralisation by a capacitor. The input balun possesses a 2:1 turn ratio, while the inter-stage transformer has a 1:1 turn ratio for maximising the output voltage swing to extend the linear region to ensure saturation at the output stage first. To achieve high-level symmetry and minimise the parasitic effects, an improved AC ground of differential topology and Metal-Oxide-Metal (MOM) capacitors have been used. The quiescent current (IQ) current-combining method sums up the two quadrature signals at the output combiner of two 1:2 turn ratio transformers for maximising output power. The design also achieved a high output power of 13.4 dBm in a phase-controlled transmitter at this high frequency of 95 GHz. However, due to the use of IQ current combining, it is difficult to show the optimum load for each of the PAs, which resulted in varying circuit performances in each of the quadrants. This effect needs to be mitigated by careful layout design [50].

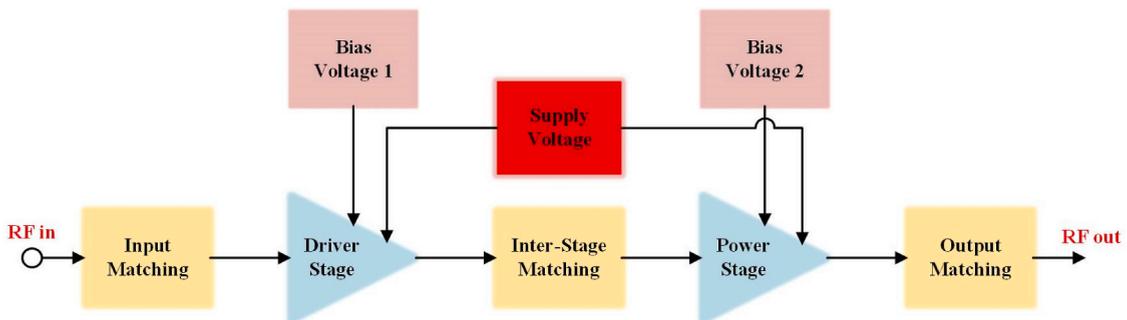

**Fig. 6.** Block diagram of a typical CMOS PA.

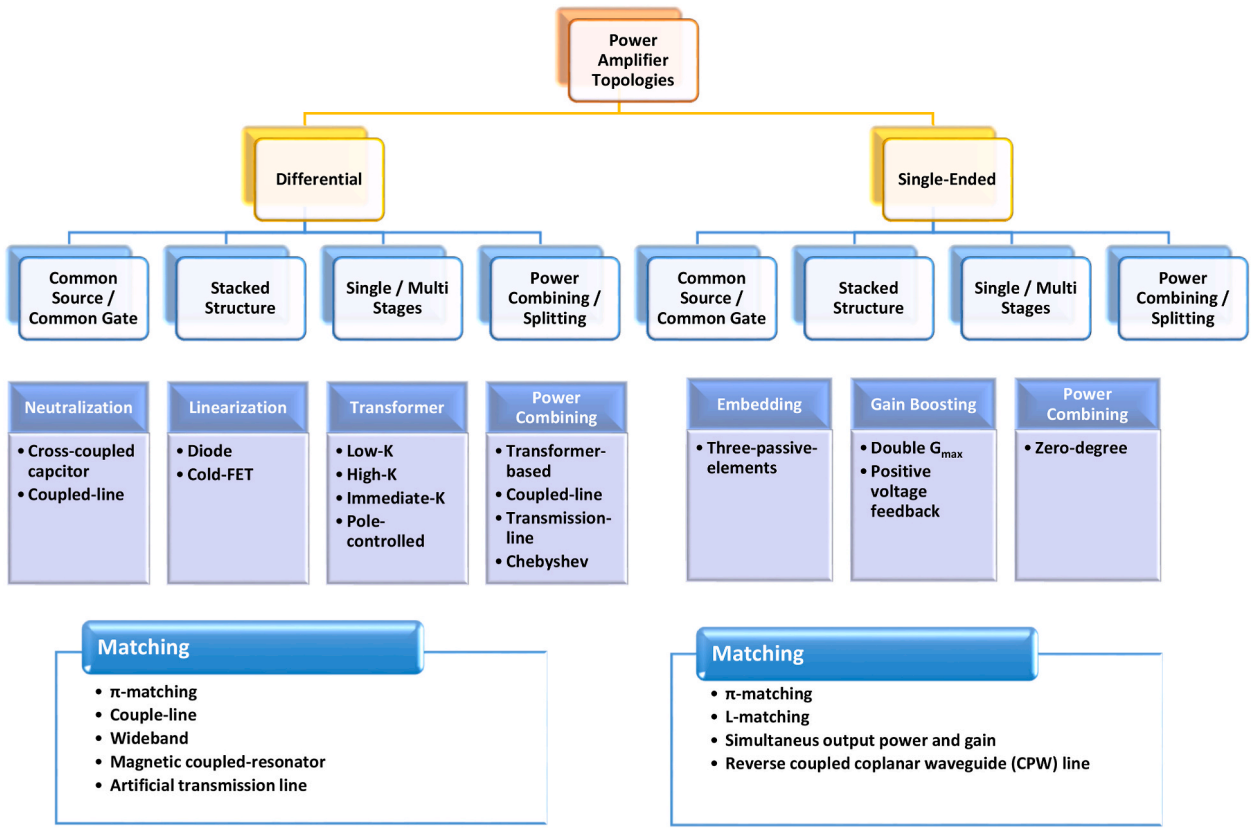

**Fig. 7.** Topologies and circuit techniques available for designing a sub-THz CMOS PA.

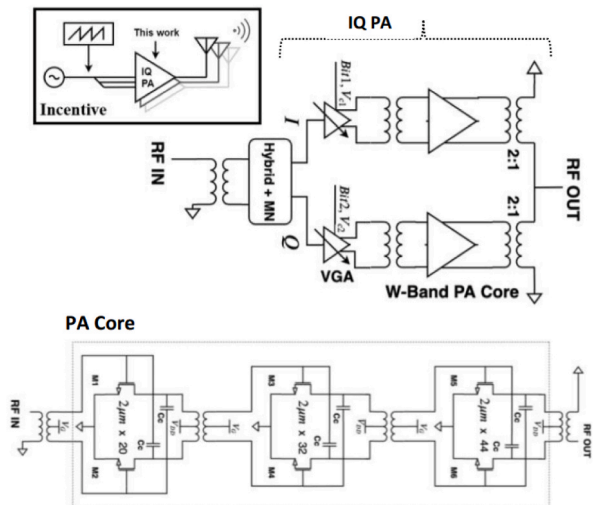

**Fig. 8.** IQ current-combining with differential Class A PA [50].

### 4.2. 128-To-1 power combining with coupled-line based PA

The PA architecture proposed in Fig. 9 by Zhu et al. [29] introduces a massive-scale power combining technique to attain high power output. Centre-fed coupled-line (CFCL) and side-fed coupled-line (SFCL) based power-combine schemes are the two fundamental units of the power combiner to effectively combine the differential signals received from the centre or the side. CFCL and SFCL can offer an inherent low-loss differential-to-single-ended power combining. In each sub-PA array, an 8-way 3 cascaded CS driver stage and a CS PA stage have been implemented. CL-based neutralisation technique was introduced with its capability to introduce an additional magnetic coupling for an efficient dynamic distribution of gate and drain voltage and to improve the output power. This technique has been included in both the output power network as well as the sub-PA. Thus far, this proposed architecture achieved the highest output power of 32.1 dBm, working at 95–300 GHz spectrum range amongst CMOS PA. However, although this design has achieved a gain of more than 20 dBm, due to the existence of multiple channels and multiple stages, it also occupies a relatively large area compared with other similar designs [29].

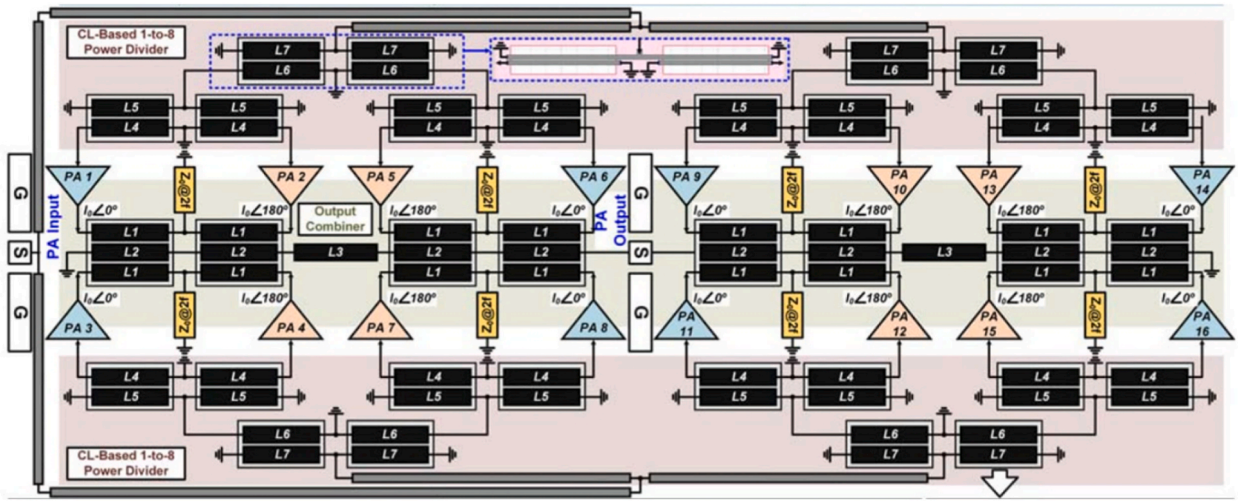

**Fig. 9.** 128-to-1 power combining with coupled-line-based PA [29].

### 4.3. 2-stages triple-stacked-FET PA

The PA topology presented by Kim et al. [51], as shown in Fig. 10, is based on the stacked field effect transistor. To counterbalance the phase of the impedance between the stacked nodes, four different combinations of parallel or series inductors have been used to analyse the stacking efficiency. Since it is a shunt/series configuration for a high quality (Q) factor and compact layout, the lowest inductance is required. Neutralising capacitors have been added to the common source pairs, to enhance stability and gain. Concurrently, to improve the power gain and reduce the gate resistance, this structure adopts a custom layout on transistor unit. The use of shunt capacitors and shunt inductors to generate second-harmonic short circuits also improves the performance of the design to some extent. Although the stacked-FET topology is used, the total output power deteriorates. Moreover, the influence of the finite Q-factor is different amongst the series and the shunt inductances. The low Q-factor of the inductor also needs to be considered due to reduction of stacking efficiency [51].

### 4.4. 4-stages CS transformer-based current combining with diode linearisation PA

The PA schematic, in Fig. 11, proposed by Son et al. [52], compromise four common source differential amplifier stage and an on-chip four-way transformer-based current combining. Furthermore, the transformer is designed with two metal layers for primary as well as secondary inductors to optimise the impedance power matching of the power stages. A capacitive cross-coupling neutralisation method was also included between the drain and the gate of each of the opposite side transistor stages. To further improve the PA linearity, a CMOS diode linearisation is used in the last stage to charge the gate-source capacitor of the transistor, to half the RF signal when the RF input power is high. Therefore, the design can show good performances of gain (20.3 dB), output power (15.2 dBm), OP1 dB (12.5 dBm) and stability. However, despite the use of capacitive cross-coupling neutralisation between the drain and the gate of each stage, the nonlinear $C_{gd}$ of the power transistor remains one of the main AM-PM distortion sources. When compared with other designs of the same type, the overall performance is not very prominent [52].

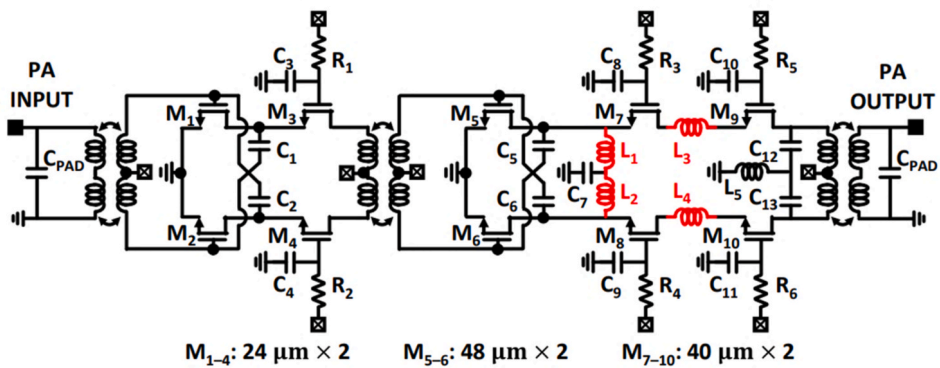

**Fig. 10.** Schematic of 2 stages triple-stacked-FET PA [51].

### 4.5. 2-stages BEOL Skip-Layer FinTech PA

The PA schematic, as shown in Fig. 12, presented by Yu et al. [53], is an F-band PA designed with a new back-end line (BEOL) using Intel 16 nm technology. In this design, the skip-layer is introduced in the PA transistor array, to reduce the parasitic loss of the thin metal layer by forming a direct connection with the thick metal layer, reducing the performance degradation through BEOL and improving the performance of PA. In high-frequency design, connections between adjacent thick metal layers offer flexibility. The resistance and the capacitance are also reduced by skipping the layer vias, thereby increasing output power, gain and PAE to 11.8 dBm, 17.1 dB and 23.8 %, respectively. The most significant advantage of this structure is that a only a few stages can be deployed for providing the required gain and can be tightly integrated into the phased array or waveguide-based transceiver. Even with neutralisation transistors, large voltage swings at the gate and the drain nodes still have an impact on the capacitance of the transistors as well as reduce the PA linearity. The metal-oxide-metal (MOM) capacitor has been selected as the $C_n$, however, this may lead to additional performance changes [53].

### 4.6. 4-stages CS transformer-based current combining with diode linearisation optimized PA

The PA design by Son et al. [54], as shown in Fig. 13, is a D-band linearised on-chip PA topology. The PA design shares the same design method as the author's previous one [52] with better optimisation. This structure neutralises each stage by a differential CS amplifier stage with a cross-coupled capacitor to mitigate the parasitic. On this basis, to improve the output power performance, direct combining with current combining transformer is also used. At the same time, the diode connection method is used to linearise the transistor, which improves the 1-dB compression point (P1dB). The power stage adopts a current combination with high output power. The high current in the secondary coil leads to a higher sensitivity towards parasitic resistance, which caused the current handling capability of the secondary coil to be low. The channel symmetry structure of this combining alleviates the deterioration of the transformer caused by the mismatch of amplitude and phase. The low reverse isolation is significantly frequency-dependent because of the high parasitic gate-drain capacitance ($C_{gd}$). This feedback leads to the deterioration of gain and stability. Due to the insufficient input power, this D-band commercial PA was deployed in a setup comprising large signal [54].

### 4.7. Pole-controlled transformer based interstage matching network PA

The PA scheme, presented in Fig. 14, proposed by Son et al. [55], consists of a three-stage differential amplifier: while the first two are driver stages with an inter-stage matching network (IMN), the last one is a power stage. The first driver stage focuses on wideband matching, the second stage focuses on a conjugate matching network to improve gain and the final stage focuses on power matching for high output power. While the transformer' primary coil will remain fixed, the secondary coil will be adjusted to control the pole. An IMN based on high-K transformer can provide with wideband characteristics. The design also used MOM capacitive neutralisation in each stage and a cold-FET lineariser in the final stage for achieving gain and linearity enhancement. Using an IMN based on high-K transformer can produce a wide range of weight characteristics, however, it may also result in an increased in-band ripple. Furthermore, using a low K induce will lead to higher losses. In addition, due to the lack of input power supply, D-band commercial PA has been added to the D-band large signal setting [55].

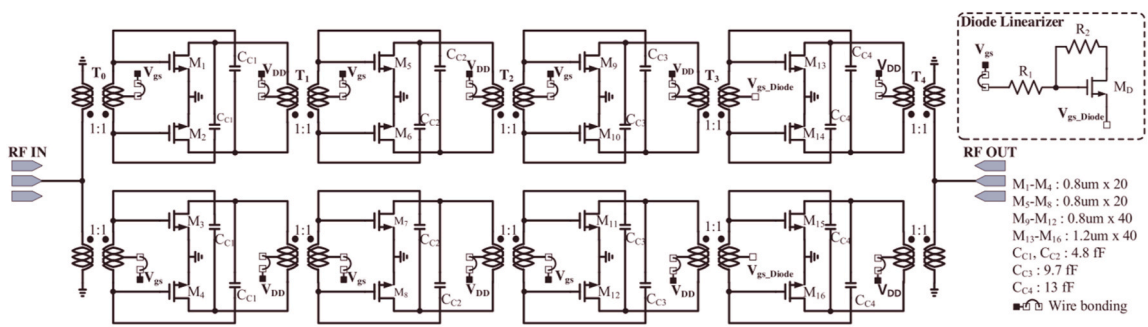

**Fig. 11.** Schematic of 4 Stages CS Transformer-based current combining with diode linearisation PA [52].

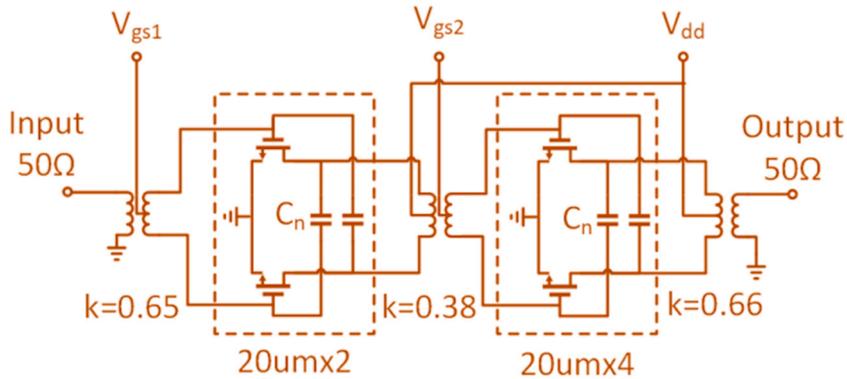

**Fig. 12.** Schematic of the 2 stages BEOL Skip-Layer FinTech PA [53].

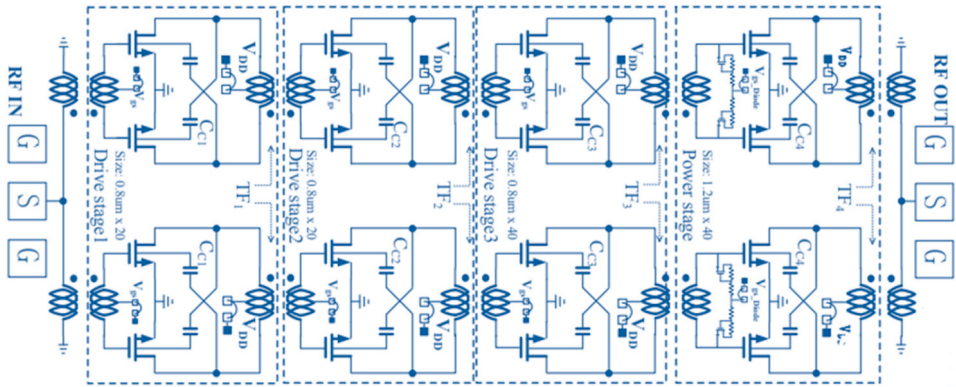

**Fig. 13.** Schematic of linearised differential CS on-chip PA topology [54].

### 4.8. 3-stages differential CS with cold-FET linearisation PA

The PA design, by Lee et al. [56], as shown in Fig. 15, possesses a power stage to achieve high output power and two driver stages to attain amplification. The Cross-coupled capacitors have been used to minimise the effect of feedback at each level. An on-chip transformer has been utilised to create a matching network that connects each stage's differential common-source amplifier. To achieve high output power levels, current combining has been used at the power stage. Additionally, a cold-FET linearisation was utilised in the final stage to improve linearity. Because of the frequency shift of the Quadrature Injection Locked Oscillator (QILO) tuning range, the measured conversion gain and the 3 dB bandwidth are significantly different. However, the 1 dB compression point of this design is relatively small [56].

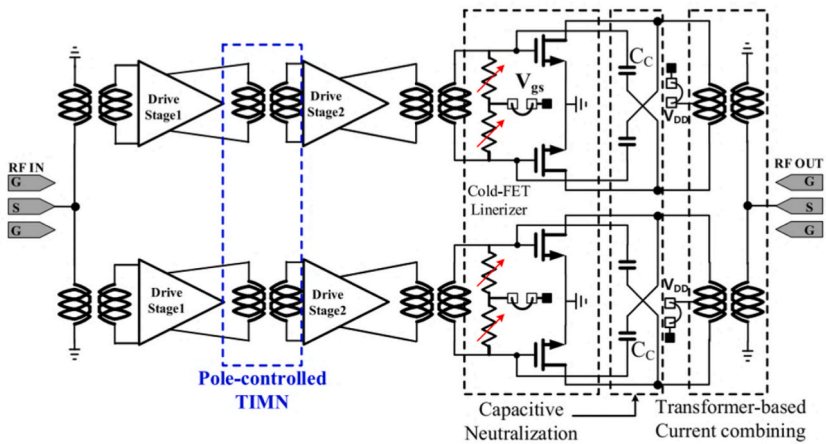

**Fig. 14.** Schematic of pole-controlled transformer-based interstage matching network PA [55].

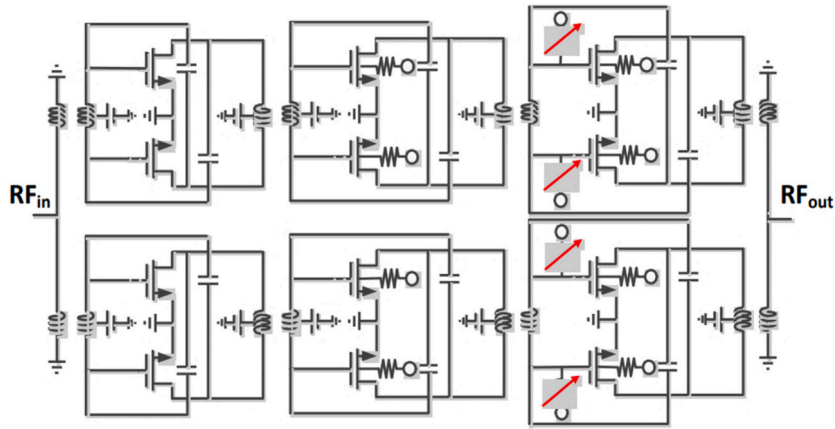

**Fig. 15.** Schematic of 3 stages differential CS with cold-FET linearisation PA [56].

### 4.9. 4-stages CS transformer-based gain-boosting Class-AB PA

The PA schematic, demonstrated in Fig. 16 [57,58], uses an efficient transformer-based matching network (TMN) design process, which optimises impedance transformation as well as passive efficiency. It refers to a highly compact D-band 4-stage PA structure and applies two different passive gain-boosting techniques to it, viz. capacitive gain-boosting and inductive gain-boosting, to achieve the optimal results of the D-band PA of the testbench. The capacitor gain-boosting technology is realised by balancing stability which can make the 4-stage PA unconditionally stable. The layout of the power transistor with a round-table further reduces the resistance of the gate and substrate, thereby increasing the $G_{max}$. At present, such PA consumes a small core integration area of 0.0265 mm$^2$ and obtains a competitive figure of merits (FoM) of 81.1, which is suitable for phased array elements in D-band communications systems. Due to the high conversion ratio, the increase of output resistance and $P_{out}$ has been limited by the added insertion loss of the balun. Meanwhile, the FD-SOI transistor is greatly affected by the operating temperature. $G_{max}$ and power efficiency also need to be traded off [57,58].

### 4.10. 3-stages transformer-based π-matching fully differential PA

The PA architecture, as shown in Fig. 17, designed by Su et al. [59], consists of a two-stage neutralised amplifier as well as a one-stage concatenated amplifier for improving the gain, isolation and the amplifier's output power through the characteristics of various amplifiers. A π-matching networks is utilised to enhance the gain and power efficiency by inserting the transmission line between the common gate and the common source transistors. In order to ensure the reasonable length of the transmission while maintaining high gain, the capacitance at the gate of the cascade code structure is also changed. The isolation (S12) of the whole D-band is better than −77dB due to the use of a neutralisation amplifier and a cascade amplifier. The amplifier adds the On-Off Keying (OOK) signal to the transistor' gate. The drain as well as the source are connected to the input end of the neutralisation amplifier. Therefore, a complete OOK modulation with a 20Gbps transmission rate can be obtained in this PA. Due to miller effect, it is difficult to achieve high gain using cascaded code amplifiers in the millimetre band. Gain and power efficiency are improved in the π-matching network, but a slight linear decrease is achieved and a well-balanced layout cannot be achieved [59].

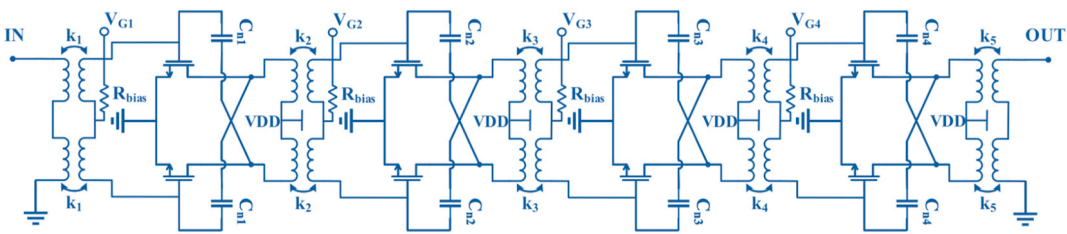

**Fig. 16.** Schematic of 4 stages CS transformer-based gain-boosting Class-AB PA [57,58].

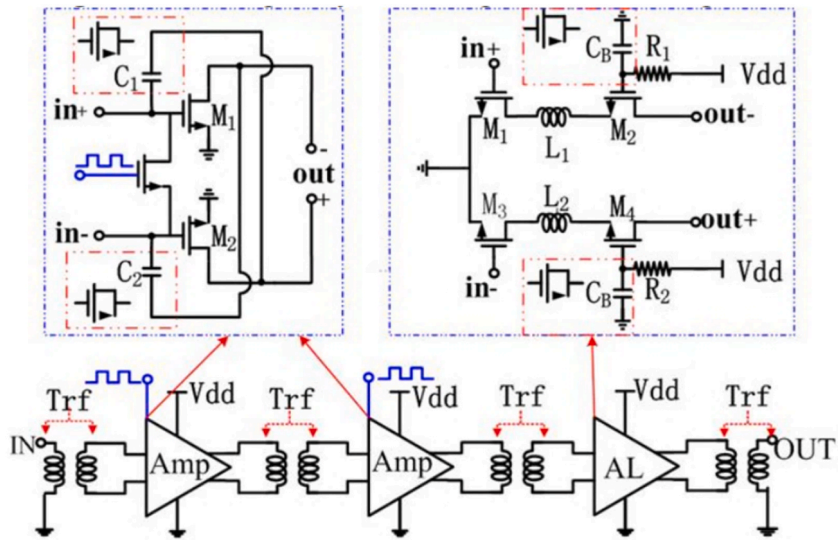

**Fig. 17.** Schematic of 3 stages transformer-based π-matching fully differential PA [59].

### 4.11. 4-stages differential neutralised 8-way CS power combined PA

The PA circuit illustrated by Li et al. [60,61] incorporates four-stage differential PA with simple and fewer passive components to achieve a high-performance PA, as shown in Fig. 18. In Fig. 18(a) the first two stages work as the amplifier stage, the third and the fourth stages are the driver stage and the power stage, respectively. The capacitor on the PA's output-matching network, with a 1:1 coupled balun of the T-combiner, is removed to cut down the losses at output-matching network as well as the combiner in Fig. 18(b). The design of a high coupling factor, in order to minimise output-matching loss, improves the output power as well as efficiency. Moreover, capacitive neutralisation using NMOS neutralised differential transistor pair is adopted in each stage to increase the maximum available gain and overall efficiency. Although this design shows excellent performance compared with the most advanced CMOS, there is still room for improvement in the area occupied [60,61].

### 4.12. 3-stages dynamic bias scaling PA

The PA structure designed by Rahimi et al. [62] describes a dynamic three-stage offset difference scaling, as expressed in Fig. 19 to adapt the PA's input power and enhance the overall Power-added efficiency (PAE) within the linear range. In CMOS Fully Depleted Silicon On Insulator (FD-SOI) technology, it uses the back-gate terminal for optimising the power consumption at all the levels. To upsurge the PA stage's current consumption with adaptive input power, the dynamic bias technology in Fig. 19(a), based on the characteristics of the back-gate terminal within the FD-SOI technology, is used to automatically tune the back-gate bias of the transistor. To improve the bandwidth, this design also uses the coupling resonator bandwidth expansion technology in Fig. 19(b). The application of the adaptive bias method in each CS stage also conically scales up the power consumption from the input to the output. The efficiency improvement technology used in this design effectively improves PAE and PBO at high frequencies and the bandwidth also has excellent performance compared with the same type. However, it still lacks competitiveness in terms of the output power and the gain [62].

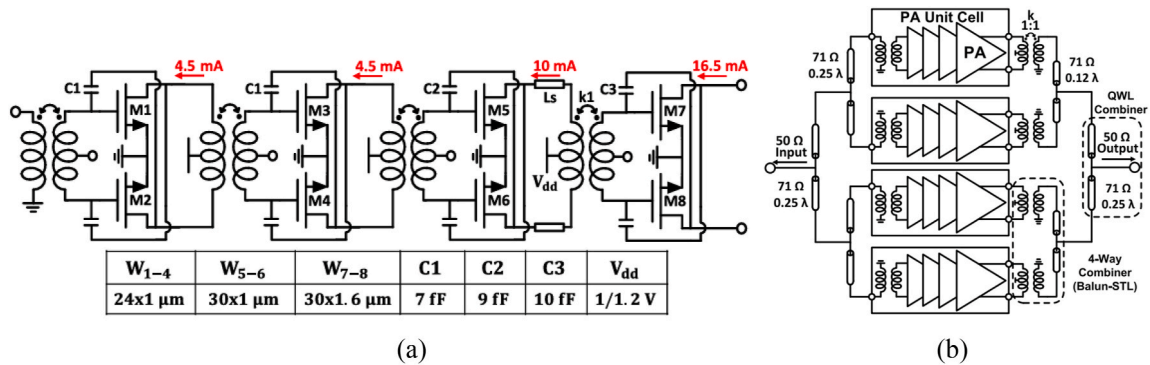

|         | (a)                                     |         | (b)                                    |

**Fig. 18.** Schematic of (a) 4 stages differential neutralised and (b) 8-way CS power combined PA [60,61].

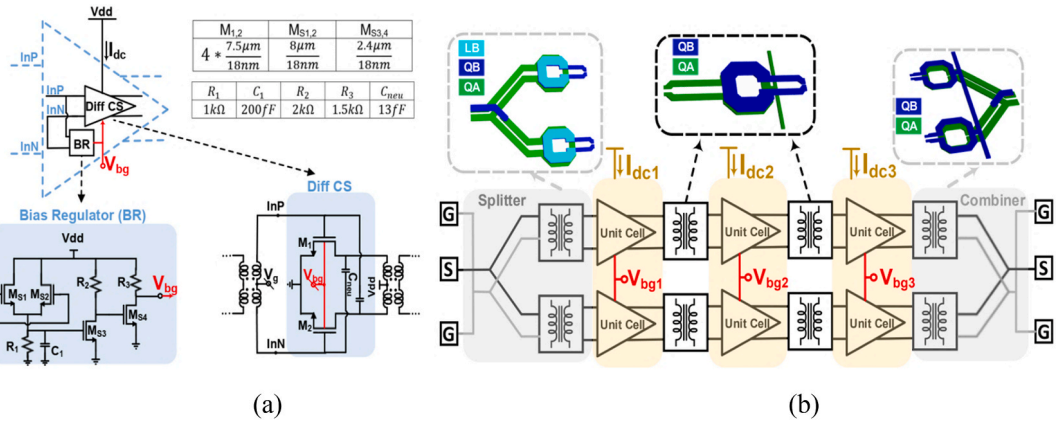

**Fig. 19.** Schematic of (a) unit cell and (b) 3 stage dynamic bias scaling PA [62].

### 4.13. 3-stages CS transformer-based with artificial transmission lines pseudo-differential PA

PA architecture featured by Zhang et al. [63] shows the topology of a 3-stage CS pseudo-differential PA with a two-channel power combination, as shown in Fig. 20. Each branch of the design adopts a 3-stage CS pseudo-differential scheme that can improve the maximum stable gain (MSG) as well as the stability. Double-ended gate contacts are also used to further enhance the MSG. In the selection process of the transistors, series feed (DS) transistors are also used to facilitate parallel connection. Because the single transistors are very narrow, they do not contribute much toward the overall width of the PA. In order to obtain larger operating bandwidth, the insertion loss is minimised by using symmetric magnetically coupled resonator (MCR) on the output matching network, while asymmetric MCR is used on input and interstage matching. An artificial transmission line composed of a shorter line and two parallel capacitors is employed for absorbing the GSG pad's parasitic which will cause impedance mismatch. This layout optimisation maximises MSG, stability and robustness. Although this design uses a small sized transistor to enhance the overall efficiency of the PA, it suffers from high equivalent resistance of the input impedance of greater than 100. The higher ratio of the impedance transform in the input matching, possesses challenges on design complexity and increase of insertion loss in the input matching network [63].

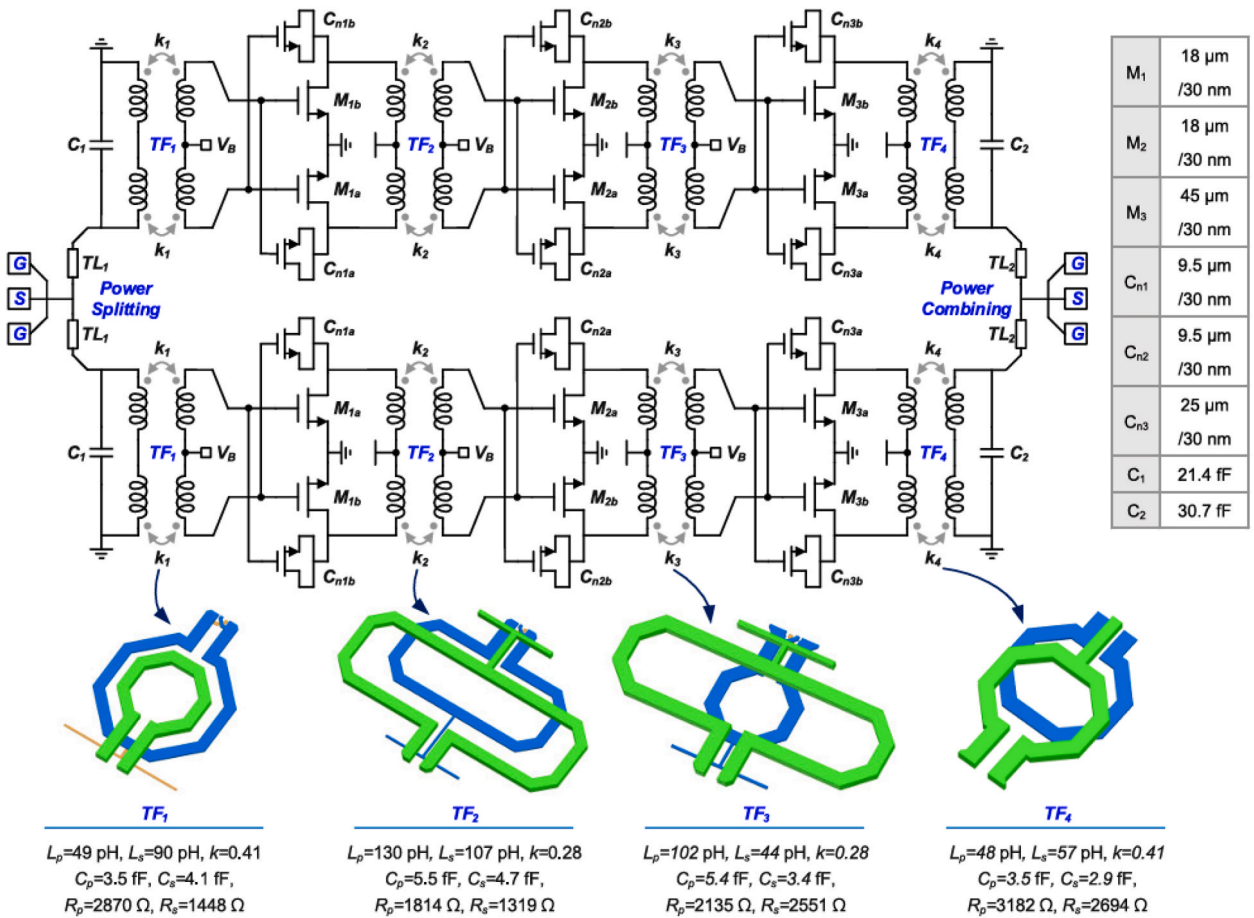

**Fig. 20.** Schematic of three-stage CS pseudo-differential PA with a two-channel power combination PA [63].

### 4.14. 3-stages CS transformer-based differential small cell transistor PA

The PA circuit, shown in Fig. 21 by Philippe et al. [64], uses a combination of small, tiltable transistors with 6 fins and 15 fingers connected to a metal layer at the top to produce larger transistors to avoid a decrease in millimetre-wave performance because of the increased parasitic capacitance as well as interconnect resistance of the lower metal and the gate electrodes. On this basis, a cross-coupled capacitor is also used to compensate for the feedback effect of the parasitic capacitance. This structure improves the transistor's stability and maximum available gain. The use of small cell transistors allows the central fin and finger have enough space to improve heat dissipation, so that it is less impacted by self-heating. The coupled output turns use two parallel inductor turns to reduce the output matching loss, and thus augmenting the output power to 15 dBm and PAE by 12.8 % of the power amplifier. This PA design is also the first to use 16 nm Fin Field-Effect Transistor (FinFET) CMOS in the sub-THz frequency band. However, the total input and output capacitance of the design limits the PA's maximum size of the power stage. The output capacitor is resonantly output by the inductor to provide with matching, but the inductance value declines as the capacitance and frequency increase, and further reducing the inductance will result in a significant increase in insertion loss [64].

### 4.15. 4-way Chebyshev artificial-transmission-line based matching wideband PA

The PA schematic presented by Zhang et al. [65], as shown in Fig. 22, is able to design a wideband PA with a steady performance. The schematic matching network is based on the magnetic coupled-resonator networks realised by a single low-k transformer in Fig. 22 (a). NMOS transistors are used for the neutralisation of the cross-coupled capacitors. The design uses an open stub in order to achieve the shunt capacitor for design feasibility. To improve the output power, a Chebyshev-type power combiner is designed from a quarter-wave length artificial transmission line without degrading the wideband operations in Fig. 22(b). The design demonstrates high performance which is suitable for a high-resolution radar, considering its wide bandwidth of 28 GHz at such a high frequency. However, in the output matching network, the choice of metal width needs to be weighed between its series resistance and parasitic capacitance. Because the on-chip device model is inaccurate at high frequencies, the difference becomes even greater [65].

### 4.16. 4-stages reusable unit-cell sandwich structure PA

The PA topology shown in Fig. 23(a), by Tang et al. [66,67], is based on a stacked field effect transistor and transformer matching network. The first three driver stages in Fig. 23(b) utilised cross-coupled capacitors to maintain stability. The transistor arrays used therein adopt a reusable unit-cell layout, which enables multistage PAs to maintain uniform performance with minimal parasitic. Such structures can provide output power without further reducing the gain. The passive gain enhancement technique is also used to improve the performance of the PA. While in the last stage in Fig. 23(c), it adopts 2-stacked-FET for high supply power and output power. At the same time, a sandwich structure is formed by the two metal layers within the secondary coil, which makes the insertion loss of the inter-stage matching network extremely low. The cell layout strategy also allows this structure to build a large transistor array to reduce parasitism. However, the PA in this design must function at frequencies close to $f_{max}/2$. This limitation of the design inevitably leads to low power gain (Gp) [66,67].

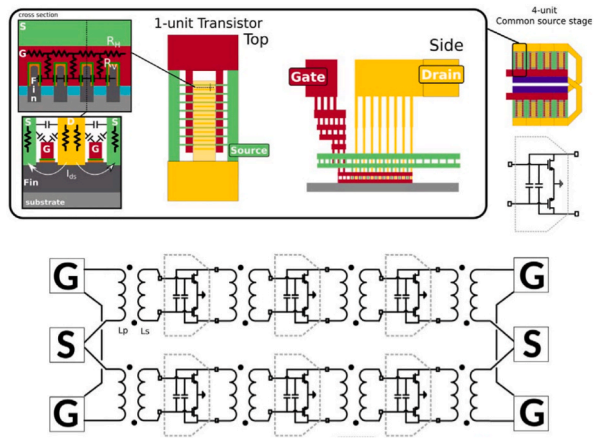

**Fig. 21.** Layout of the unit cell in amplifier stage and schematic of 3 Stages CS transformer-based differential small cell transistor PA [64].

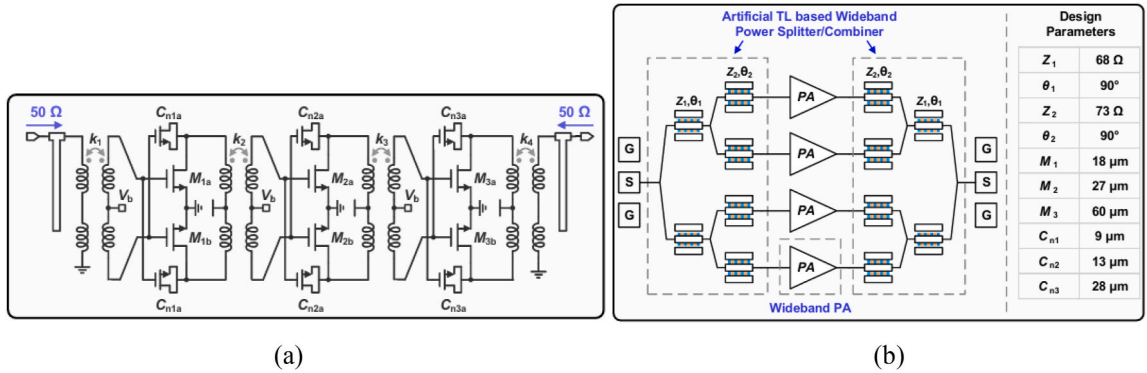

(a)                                                                 (b)

**Fig. 22.** Schematic of (a) single channel PA and (b) 4-way Chebyshev artificial-transmission-line based matching wideband PA [65].

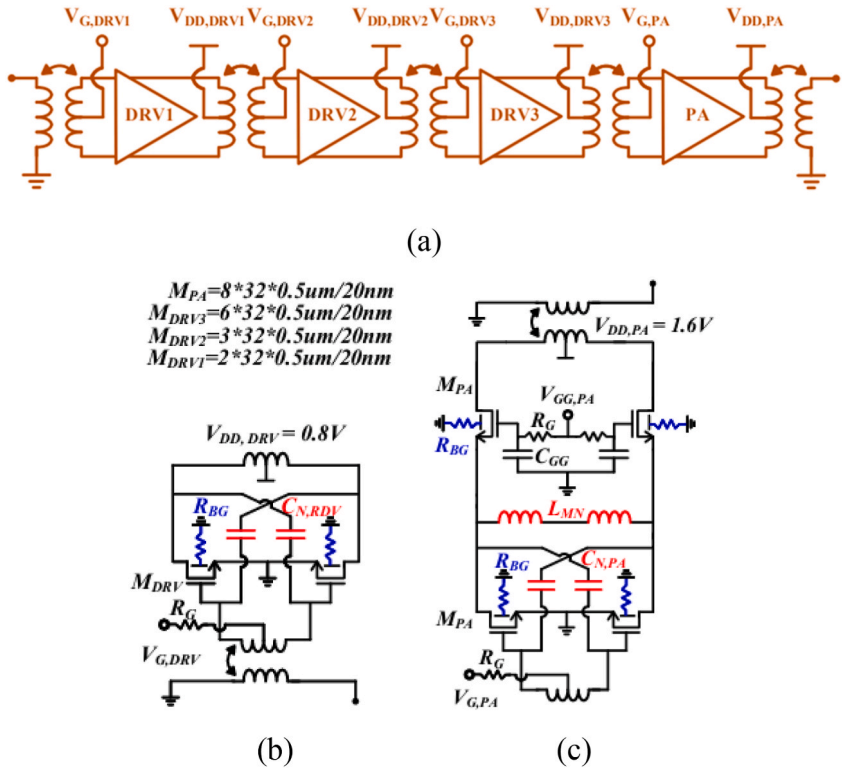

**Fig. 23.** Schematic of (a) 4-stage PA, (b) driver stage and (c) output stage [66,67].

### 4.17. 4-way symmetric balun based power combiner PA

The PA circuit designed by Yamazaki et al. [68], as shown in Fig. 24, uses a balun-based power combiner in order to raise the output power. The problem with layout symmetry leads to inaccurate impedance matching at high frequencies. An improved symmetry can be realised by grounding the secondary coil. The centre of the balun structure has to take out to allow subsequence circuit blocks to be connected. A lower layer of metals is utilised for extracting the output power to the outside due to the high frequency which makes it challenging to retain symmetry. The more parallel PA stages are included, the smaller the deviation from the optimal impedance; the performance will also significantly decrease. The PA gain in this design is low, which needs to be improved through cascading multiple amplifier stages in each PA [68].

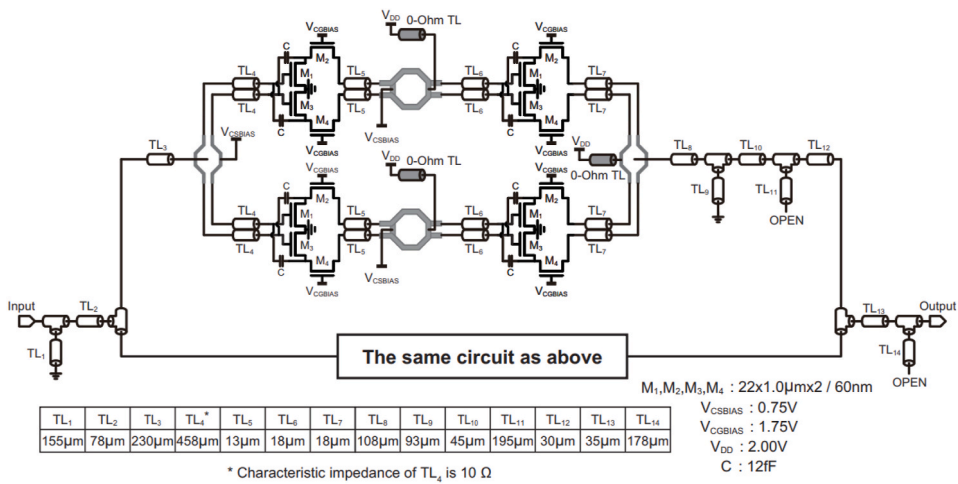

**Fig. 24.** Schematic of symmetric balun-based power combiner PA [68].

### 4.18. 3-stages cascode transformer-based GSG-pad optimized PA

The PA design by Simic et al. [69], as shown in Fig. 25, shows a two-way three cascaded stages pseudo-differential pair with cross-coupled custom-made parallel plate capacitors. The custom capacitor structure will have better accuracy when conducting the modelling since the majority electrical field will be beamed in a vertical direction. Matching is conducted using sacked transformers associated with additional transmission lines to allow improved coupling with inductors, compact design and easy routing. The Ground-Signal-Ground (GSG) pad's geometry has also been altered to solve the imbalance problem caused by the deviation of the ideal load impedance from the output stage. The major flaw of this PA design, operating in the D-band high frequency, is the volume effect's uncertainty. In addition, the cascade code transistor gate nodes also cause stability problems. The power combination method increases the area of the chip and the complexity of the circuit [69].

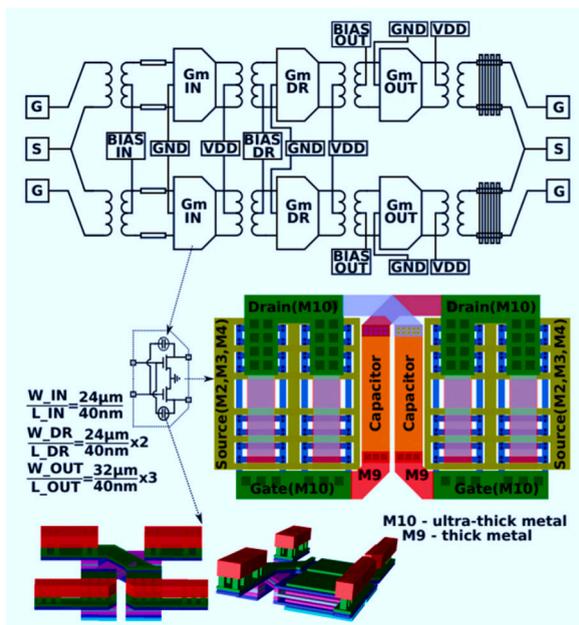

**Fig. 25.** Schematic of 3 Stages Cascode Transformer-based custom-made parallel plate capacitors PA [69].

### 4.19. 3-stage fully differential CS-cascode transformer-based PA

The PA circuit described by Gao et al. [70], as shown in Fig. 26, combines the common-source (CS) amplifier with the cascode amplifier technologies to attain high outpower power and gain. While the opening two stages are built from a common-source structure, the last one is configured with a cascode stage. Cross-coupled neutralisation capacitors were used in all the stages to retain its gain and stability. Furthermore, two thick metal layers on top of the bulk process technology were used to build the transformer and reduce the parasitic. This could reduce the substrate loss in the differential mode. Due to the usage of cascode structure, the voltage supply will be higher than the nominal amount provided for the CS stage which will cause design complexity, as a separate regulator will be required [70].

### 4.20. Gmax based simultaneous gain and output power matching PA

The PA architecture featured by Park et al. [71] adapts the $G_{max}$ concept of simultaneous gain as well as output power matching techniques, as shown in Fig. 27, which efficiently reduces the number of amplification stages to only two: power amplification and driver amplification. The gain and output power matching are achieved from an embedding network based on 3-passive elements, to enable the point of identical matching for both small as well as large signals. This method adopts both power and driver amplification stages. The LC resonance concept is able to achieve the $G_{max}$-based gain-boosting. Thus, this architecture can provide an extra degree of freedom, maximise small-signal gain to 17.5 dB, large-signal output power to 10.4 dBm and output power 1 dB compression point ($OP_{1dB}$) to 5.3 dBm. However, the inaccuracy of MOSFET, passive elements and substrate coupling easily degrade the gain and bandwidth results due to high operating frequencies [71].

### 4.21. 4-stage pseudo-differential neutralised CS cascaded PA

The PA circuit, shown in Fig. 28 by Hamani et al. [72], is composed of 4 stages of cross-coupled capacitive neutralised pseudo-differential CS amplification. The CS transistor is cascaded in two ways by a transformer that is configured as baluns. There is also an inter-stage transformer between the inputs and the outputs to maintain the differential mode. To reduce the parasitic effect and the resistive metallic losses, the transformers are implemented in the thickest conductors in BEOL. Open stub matching has been placed at the output port to improve the matching. Thus, this circuit can cover a wider bandwidth of 17.5 GHz with a competitive power consumption of 92 mW as well as an area occupation of 0.04 mm². The problem of PA operated at high frequency is still clearly observed as the gain, output power, and $OP_{1dBm}$ degrade because of the transistor model` limited accuracy [72].

### 4.22. 5-stages single-ended CS with coplanar waveguide (CPW) based positive voltage feedback PA

The schematic, highlighted in Fig. 29 by Zhang et al. [73], has introduced positive voltage feedback using a pair of coplanar waveguides (CPW) based inductors on a single-ended 5 CS stages PA. The transistor size scale of stages 1–3:4:5 is 1:2:4. Stages 1–3 adopt a set of two coupled grounded-CPW, while stages 4–5 have another set of two reverse coupled CPW lines for improving gain boosting by tackling the frequency characteristic deterioration because of the increasing transistor scale. The matching network of this schematic uses a CPW line for enhancing shielding and reducing resistive loss as well as a π-type network to achieve flat gain frequency response. However, the CPW structure will cause inferior isolation between the source and the gate, eventually limiting the output power [73].

### 4.23. 5-stages differential coupled-transmission lines staggered wideband PA

The PA schematic put forwarded by Liu et al. [74], as demonstrated in Fig. 30(a), consists of five stages of differential pair with cross-coupled capacitors (Fig. 30 (b)) and matched with coupled transmission lines (CTL). CTL, in Fig. 30(c), is able to perform higher quality factor (Q) with a simplified model and convenient wideband circuit design. The separated peaks of the impedance curve can move apart from each other and are matched with the wide bandwidth. The CTL high-order network is furthered with a staggered inter-stage wideband matching technique which can stagger all the matching frequencies from different stages. This could eventually provide stable performance in a wideband operation at 32.8 GHz and the wideband flatness can be simply modified with the CTL structure. The inaccurate models of the transistor at the high operating frequency and high emphasising of linear gain by increasing the stages without any power combining lead to the low output power of 3 dBm as well as negative $OP_{1dBm}$ of −4.3 dBm [74].

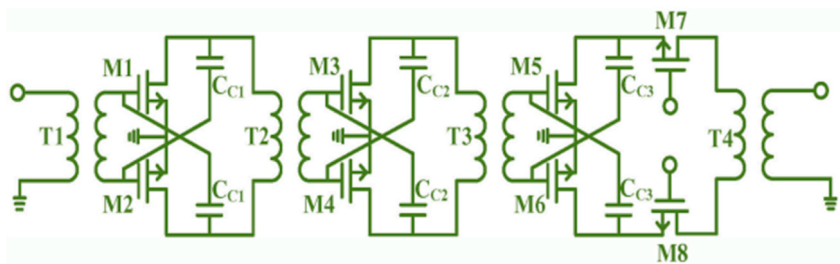

**Fig. 26.** Schematic of 3 Stage fully differential CS-cascode transformer-based PA [70].

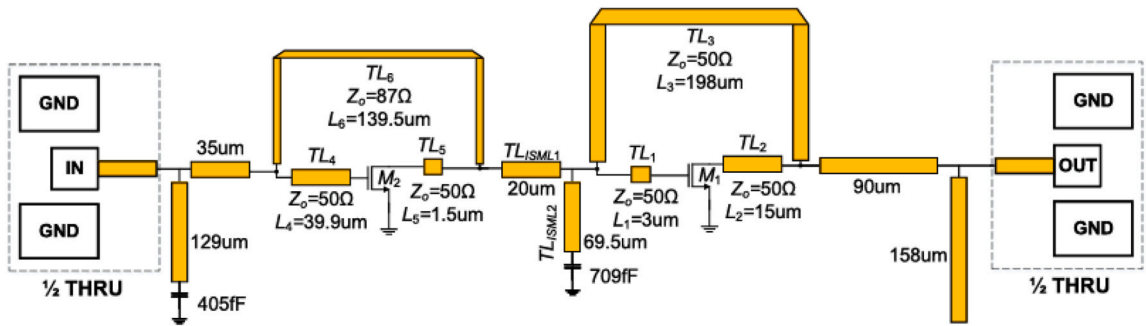

**Fig. 27.** Schematic of $G_{max}$-based PA of simultaneous gain as well as output power matching [71].

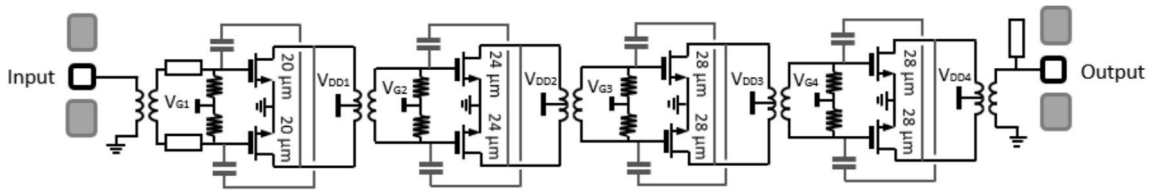

**Fig. 28.** Schematic of 4-stage pseudo-differential neutralised CS cascaded PA [72].

### 4.24. 8-stages embedded matched-cascode Amp-cell PA topology PA

The PA circuit design, demonstrated in Fig. 31 by Bameri et al. [75], shows the structure of an embedded matched-cascode amplification unit. This structure increases the supply voltage and $P_{sat}$ without sacrificing power gain. To overcome the matching loss and increase the power gain and total $P_{sat}$, embedding is used around the amplifier unit. The design adds a matching network in-between the common source and the common gate transistors to improve the output power as well as the power gain of the amplifier unit. This matching increases the power gain performance and uniformly distributes the voltage swings on the transistor nodes, allowing them to simultaneously enter the nonlinear region. The size of the transistors used was reduced in the early stages to improve gain as well as efficiency. All the amplifier units are harmonised with their intricate conjugate input/output impedances in order to maintain their gain. The necessary techniques such as cascade structure, multi-stages and power splitting to improve the gain and saturated power for operating at the high frequency required higher supply voltage which caused a drawback of high power dissipation [75].

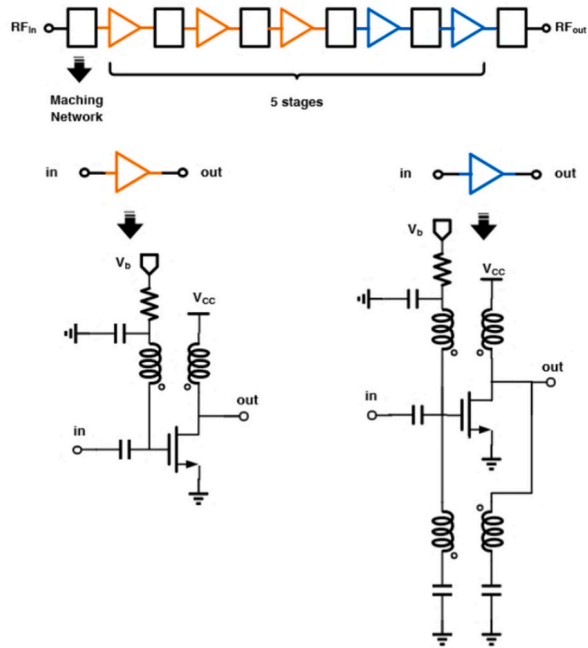

**Fig. 29.** Schematic of 5-Stage Single-Ended CS with Coplanar Waveguide (CPW) based PA [73].

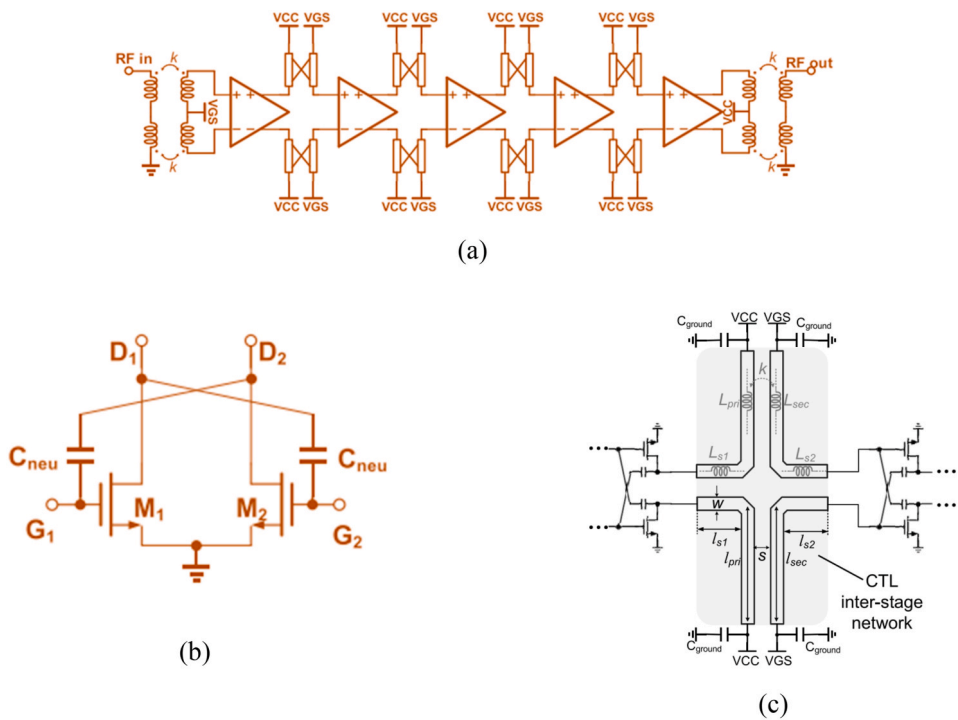

**Fig. 30.** Schematic of (a) overall PA schematic, (b) cross-coupled capacitor, and (c) coupled transmission-line network [74].

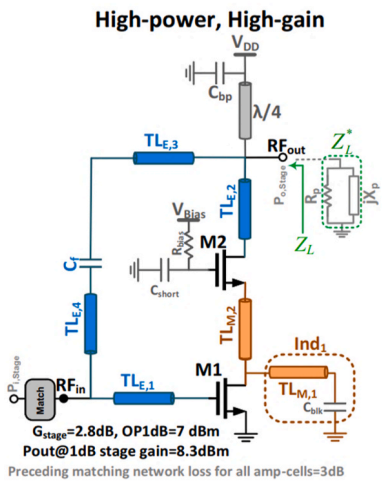

**Fig. 31.** Schematic of embedded matched-cascode amplification unit [75].

### 4.25. 8-stages pseudo-differential CS transformer-based with MIM capacitor cross-coupling PA

The PA structure, as shown in Fig. 32, proposed by Atar et al. [76], has four parallel chains, eight gain stages, and a new input and output series combination transformer. Such structures can solve the contradiction between the use of small-sized devices and the limitation of maximum current handling capacity due to the inverse ratio of $f_{max}$ to device size. For layout efficiency approach, in the input-matching inductor, an active device is instantiated. By driving the transformer in an interleaved inverse phase manner, an in-phase is formed by the magnetic field within the neighbouring transformers, causing every two adjoining output ports to be out of phase. Thus, the possibility of all the output ports of the voltage divider to be in the same phase can be eliminated, as the magnetic field formed between every two adjoining transformers is offset by each phase. This is due to the fact that the adjacent transformers are cross-coupled, thereby, reducing the efficiency of the voltage divider. However, the PA suffered from large isolation loss due to the high current leakage to the bulk, which modulated the threshold voltage and eventually caused gain expansion [76].

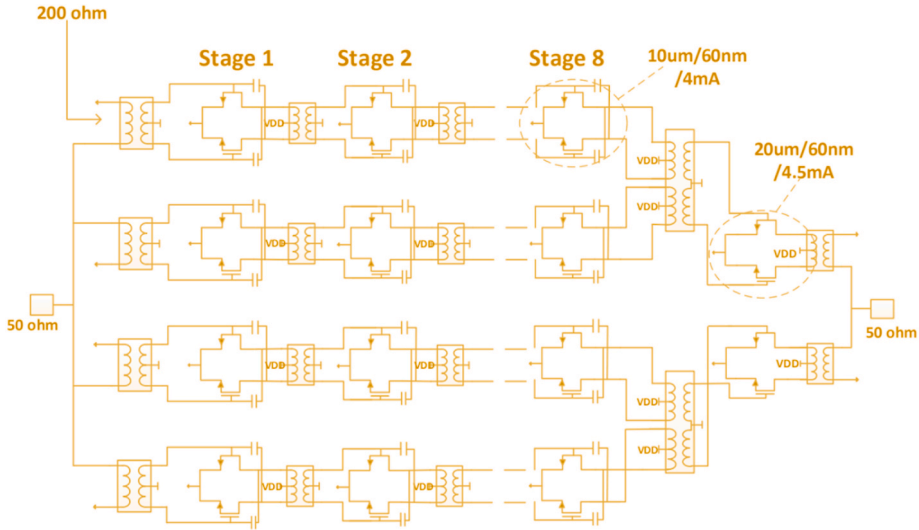

**Fig. 32.** Schematic of 8 stages pseudo-differential CS transformer-based with MIM capacitor cross-coupling PA [76].

### 4.26. 2- and 4-way transmission line-based zero-degree power combining PA

Fig. 33 shows the PA schematic, reported by Yun et al. [77], that uses zero-degree power splitter/combiner (ZDS/ZDC) based on $G_{max}$ and transmission line, for 245 GHz PA deploying 2- and 4-way power combing techniques in Fig. 33(a) and (b) respectively. This design implements each amplification phase based on $G_{max}$, by using an embedded network based on three passive elements. Concurrently, good gain and efficiency are attained by reducing the inter-stage matching networks' insertion loss as well as magnetic coupling, to ensure isolation between adjacent inter-stage matching networks (ISMs). Furthermore, in this PA, each composition stage is optimized from the final to the first stages for obtaining the highest output power of 10.5 dBm. When the $S_{12}$ increases, the stability factors degrade as a result of the undesired feedback signals originating from the substrate as well as the ground plane [77].

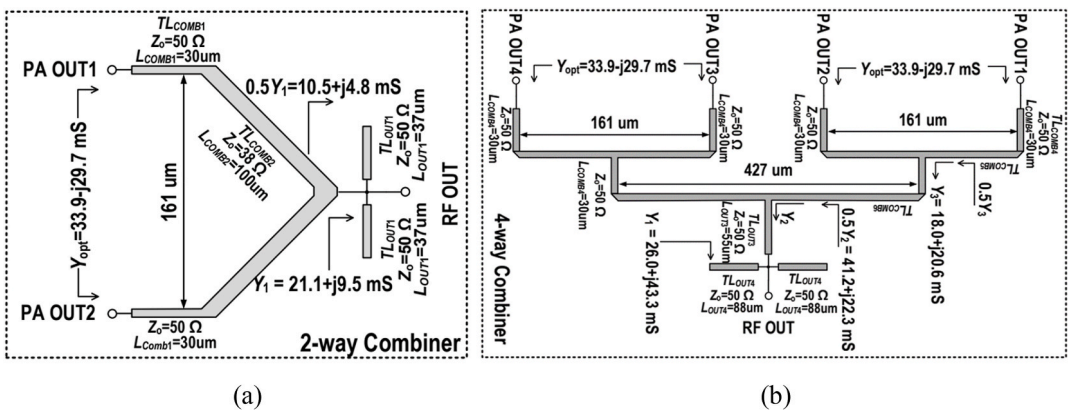

(a)                                                                                        (b)

**Fig. 33.** Schematic of zero-degree power combiner based on (a) 2-way and (b) 4-way transmission line [77].

### 4.27. 3-stages dual-shunt element-based $G_{max}$-core PA

The PA design from Yun et al. [78], as depicted in Fig. 34, is able to increase the output transistor's size by using the dual-shunt-element-based $G_{max}$-core, resulting in higher output power. To alleviate the achievable gain' limitation, the maximum achievable gain ($G_{max}$) is rather adopted. The drive stage (DS) adopts double $G_{max}$ gain enhancement technology to improve the gain and efficiency. This PA design technique is also used to overcome the size limitation of the gate-drain shunt inductor. This method is able to minimise the transmission line (TL) length of the parallel inductor but meets the physical size required for the $G_{max}$ core based on a large-size transistor, resulting in higher PA output power and efficiency. Diode-connected configuration is used to avoid design complexity and additional loss. The PA still consists of design challenges in matching the $Z_3$ TL's width due to the layout criteria limitations of metal-10 within the specified technology [78].

### 4.28. CS stage with neutralising capacitors transformer-based wide bandwidth PA

Fig. 35 illustrates a W-band broadband PA design [85]. The design fully utilises the differential CS PA along with the capacitor gain boosting technique and transformer-based matching networks. The PA is designed with class-AB operation to maintain the CS stability; a large resistor is serially embedded to the gate of the transistor in the transformer. The unit dimension of the transistors is sized with 12 fingers to provide high output power. The neutralising capacitor has also been used to improve overall gain and maintain stability. The CS gain boosting efficiently provide with a lower Q-factor for input and output impedances to allow a better wide-band matching using transformer balun. This combination can allow wideband operation while maintaining the amplifying gain across the bandwidth. However, the design can consider N-way power combining to boost the overall saturated output power and efficiency performance.

### 4.29. De-overstabilisation based four-way combined D-band PA PA

The design in Ref. [86], as shown in Fig. 36 (a), consists of a four-way combined D-band PA target on output power and efficiency performances. This work introduces a de-overstabilisation technique on top of capacitance neutralisation to boost overall PA stability at high-frequency operations. The de-overstabilisation technique takes into account the RF resistance across the matching network. In Fig. 36 (b), a distributed active transformer (DAT) power combining has been put forward to combine the two differential pairs of the PA, to efficiently reduce the ohmic noise. The DAT's compensating inductors also recompense the RF signal leakage caused by imbalance. This method can efficiently contribute to a high coupling coefficient as well as low RF resistance while increasing the output power. Instead, it sacrifices the PA gain performance and introduces the usage of area-hungry inductors.

## 5. Discussion

CMOS PA design has attained notable interest because of its high performance as well as low-cost capability. For applications such as 6G communications, PA in CMOS technology is considered to be one of the potential technologies to develop, compared to others. FCC also opened the THz spectrum band between 95 GHz and 3 THz for the research and development of 6G systems [14]. Therefore, the sub-THz frequency band can be developed for 6G communications, particularly for low-frequency operations. Table 4 highlights several potential optimising techniques for different specifications, such as BW, gain, saturated output power, PAE and OP1dB. Furthermore, different states-of-the-art CMOS PA designs published over the past eight years, along with their relevant performance parameters as well as circuit topologies operating within the sub-THz frequency from 95 GHz to 300 GHz, have been compiled in Table 5. The comparison, as presented in Table 5, contains significant PA design specifications including bandwidth, gain, saturated power, OP1dB, PAE, area consumption, power dissipation as well as FOM. Each specification in the table has been distinctly analysed to identify the three most competent designs and convenient circuit topology approaches in the sub-THz frequency range for a given design metric. That being said, the power dissipation and FOM values have been appended in the table but were not closely scrutinised due to the lack of sufficient data and calculations from non-identical equations.

In 6G wireless communications system, ensuring high data rate is very necessary due to the increased demand for connected devices and consumer needs. The wide bandwidth offered by the 6G communications system can attain multi-Gbps data rate to meet these requirements [55]. The PAs designed in Refs. [55,56,74] have been singled out for wide bandwidth design with 3 dB BW of 40 GHz, 38.5 GHz and 32.8 GHz, respectively. All of these three designs adapted different techniques to achieve wideband operations. The architecture presented in Ref. [56] utilised a resistive feedback inverter for the up-conversion mixer and degraded its conversion gain to achieve wideband operations. On the other hand, in the PA the design presented in Ref. [55], in each amplifier stage with wideband operations to solve the multistage amplifier BW degrading problem, the drive stage has wider bandwidth than the power stage, in the two-stage condition. The design from Ref. [74] utilised the coupled transmission lines (CTLs) inter-stage matching to realise wide bandwidth operations. In general, the design presented in Ref. [55] shows a better overall performance with wideband operations compared to the other two. The PA design of [56] offers the largest bandwidth but it deteriorates after being connected to the transmitter.

Capacitive parasitic effects have become a major challenge when designing PA in the high frequency of the sub-THz spectrum, especially for 6G communications applications. These will cause a serial effect towards the gain, linearity and efficiency [61]. By comparing different sub-THz PA design, [29,58], and [77] stands out as the best three, in terms of their gain performance with 33.6 dB, 32.5 dB and 28.0 dB, respectively. The PA presented in Ref. [58] uses a passive gain boosting method, with a combination of capacitor neutralisation and an inductor intermediate matching component. In addition, the architecture presented in Ref. [29] utilised coupled-line (CL) based scalable 128-to-1 power combining, dividing and neutralisation techniques to achieve the highest gain. The architecture in Ref. [77] uses zero-degree power splitters/combiners (ZDS/ZDC) based on a 2-way $G_{max}$ as well as transmission line, to avoid the coupling effect and minimise the insertion loss. Overall, the design presented in Ref. [29] demonstrated the best overall performance compared with the other two designs although it does not have the highest gain record. However, it is also obvious that this design is very complex amongst all the PA designs described in Table 5, which has eventually caused the core size to be the largest too. The design presented in Ref. [77] can operate at higher frequencies with the least complex design, smallest core area and lowest power dissipation amongst the three designs, however, it demonstrates low performance. The design presented in Ref. [58] is able to operate at higher frequencies with an integrated transmit/receive (T/R) switch but demonstrates average performance.

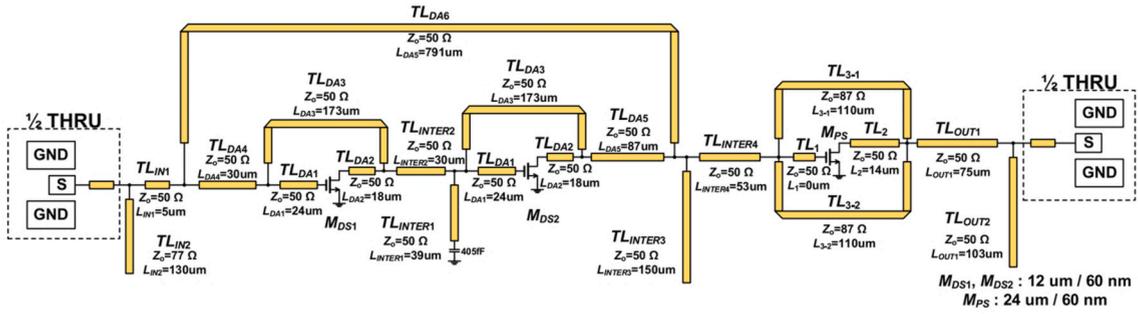

**Fig. 34.** Schematic of 3 stages dual-shunt element-based $G_{max}$-core PA [78].

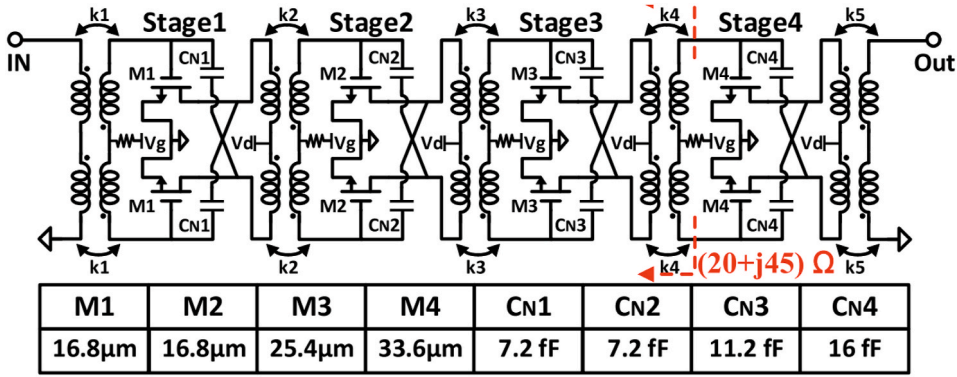

**Fig. 35.** The schematic of CS stage with neutralising capacitors transformer-based wide bandwidth PA [85].

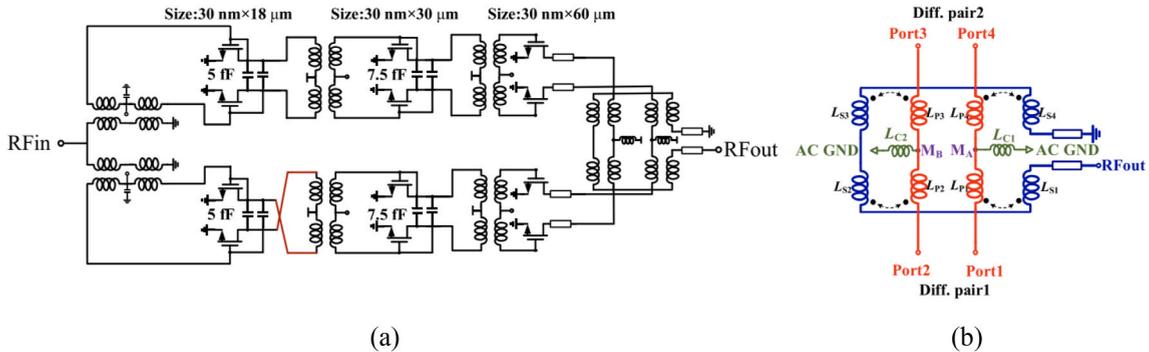

(a)           (b)

**Fig. 36.** The design of a (a) 3-stage CS PA with neutralisation capacitance and (b) differential DAT power combiner with compensating inductors [86].

The PAs designed in Refs. [29,61], and [52] are considered to be the best three in terms of their saturated output power performance with 32.1 dBm, 18.5 dBm as well as 15.2 dBm, respectively. As stated before, The PA architecture in Ref. [29] utilised coupled-line (CL) based scalable 128-to-1 power combining, dividing and neutralisation techniques to achieve the highest saturated power. The PA circuit depicted in Ref. [61] deploys an eight-way combined power integrating with four-way differential cascode amplifiers PA loss-power combining technique. Apart from that, the design presented in Ref. [52] utilises a transformer-based current combining method to increase the output power of a four-way four-stages PA. Based on the aforementioned three PA designs, it is observed that PA power and current splitting and combining methods can improve the output power. In fact, the higher the number of splitting and combining ways, the higher the output power. While [29] consists of 128 ways [61], consists of eight ways (4-ways differential), and [52] consists of four ways.

**Table 4**
Comparison between different optimising techniques for different metrics.

| Optimising Technique | BW | Gain | $P_{sat}$ | PAE | OP1dB |
|---|---|---|---|---|---|
| Resistive Feedback Inverter [56] | ✓ | | | | |
| Multistage Wideband [55] | ✓ | | | | |
| Passive Gain Boosting [58] | | ✓ | | | |
| Coupled-Line [29] | ✓ | ✓ | ✓ | | ✓ |
| Zero-Degree Power Combining [77] | | ✓ | ✓ | | |
| Integrated Power Combining [61] | | | ✓ | | |
| Transformer-Based Power Combining [52] | | | ✓ | | |
| Skip-Layer Vias & Continuous Vias [53] | | | | ✓ | |
| Second-Harmonic Short Circuit [51] | | | | ✓ | |
| $G_{max}$-based output power and gain matching [71] | | ✓ | | ✓ | |
| Transformer Coupled [61] | | | ✓ | | ✓ |
| Diode Linearisation [52] | | | ✓ | | ✓ |

The PAs presented in Refs. [51,53,71] demonstrate peak PAE of 23.8 %, 18.6 %, and 16.1 %, respectfully. The PA design depicted in Ref. [53] introduces skip-layer vias and continuous vias in the PA transistor design to reduce resistance and capacitance, eventually improving efficiency. The PA architecture in Ref. [51] inserts an extra second-harmonic short circuit on the drain of the triple-stacked transistor for optimal load-pull matching to reach high efficiency. The PA in Ref. [71] adopts simultaneous output power-as well as gain-matching based on $G_{max}$, by optimising an embedding network based on three-passive-elements to achieve high efficiency without any power-combining. Overall, the design presented in Ref. [53] is considered as the best overall design amongst these three designs with the highest PAE. Although design presented in Ref. [51] demonstrate better performance in terms of the other parameters, the supply voltage is three times higher compared with the other two designs, which eventually results in better overall performance than the others.

From Fig. 37, it is important to observe the limit of the OP1dB value to allow the PA to operate linearly. The PA designs presented in Refs. [29,61], and [52] show the highest OP1dB value of 28 dBm, 14.2 dBm and 12.5 dBm, respectively. The PA architecture of [29] uses the CL-based power combining method to allow the active load modulation behaviour to maintain the highest OP1dB. The PA circuit in Ref. [61] utilised the transformer-coupled method for a high OP1dB. On the contrary, the PA design of [52] uses the diode linearisation method to enhance the OP1dB. All of these three architectures of high OP1dB possess a similarity of having high saturated power. Therefore, the OP1dB value is highly dependent on the saturated power of the PA device, as the PA will be linear if both values are closer, as shown in Fig. 37. A proper linearisation will be required to maintain the linearity so that the PA will work in a larger area of compression point and does not go into compression and non-linear state. However, it is observed that in order to drive PA at a better Psat, adjusting transistor structure, appropriate impedance matching, parasitic and dielectric loss compensation, etc. are required.

As evident from the aforementioned discussions and comparisons, different PA designs have their own trade-offs on different parameters such as gain, saturated power, PAE, OP1dB and area consumption, to achieve the design objectives. In terms of the 6G communications, the CMOS PA has several important factors which need to be prioritised to provide with a high data rate, to exchange huge amount of information. Highly efficient wideband operation in high frequency is one of these important factors. Therefore, the PA stages plays a vital role, as they can improve the gain. However, the performances of BW and efficiency also need to be considered, as they will deteriorate when the PA stages increase. Although some of the advanced PA linearisation techniques, such as envelope tracking (ET) and digital pre-distortion (DPD), are able to solve the PA efficiency deterioration, they still possess some challenges, particularly for wide and large channel GHz modulation BW [79]. An intelligence self-sensing DPD is able to linearise the PA with an intelligence sensing approach [80]. A complexity-reduced band-limited memory method has also been proposed in Ref. [81] for wideband PA. Furthermore, another approach of design using Doherty PA (DPA) architecture has the capability to improve the PA linearity and efficiency. In another study [82], a SiGe BiCMOS process DPA working in sub-THz frequency has been proposed. There are still some concerns regarding the practical DPA usage in the sub-THz band due to the challenges faced when implementing the impedance inverter with poor passive efficiency [53]. The inverter's size reduces with increasing frequency, making it difficult to realise with high efficiency. Thus, it suffers from parasitic effects and transmission line losses due to factors such as skin effect and substrate coupling, ultimately leading to a decline in overall passive efficiency.

When designing a high-frequency device, especially in a sub-THz frequency band, it is highly important to carefully deal with the low reverse isolation due to the parasitic effects that will seriously occur when operating in high frequency on CMOS and other process nodes technology. Capacitive cross-coupled neutralisation is a commonly used method to improve the PA gain, bandwidth and stability for compensating the feedback effect due to the parasitic capacitance and to take the edge off the insertion loss. Fig. 38 shows the design interest to improve the linearity of PAs based on the designs described in Table 5. As seen from Fig. 38, twenty three (23) articles utilised capacitive cross-coupled neutralisation techniques, whereas eight (8) articles rather applied other approaches. Another four

**Table 5**
Performance statistics of different PA designs in sub-THz band. Structured.

| Ref. | Process (nm) | Technology | Topology | Power Combining | fc (GHz) | BW3dB (GHz) | Gain (dB) | Psat (dBm) | PAEpeak (%) | OP1dBm (dBm) | VDD (V) | Power (mW) | Core Area (mm2) | FoM** | Key Circuit Techniques |
|---|---|---|---|---|---|---|---|---|---|---|---|---|---|---|---|
| [50] (2021) | 65 | CMOS_Bulk | 3 Pseudo-diff. CS | No | 95 | 9 | 17 | 12 | 12.3 | 8 | 1.2 | 66 | 0.025 | – | 0.5 V biasing, stacked inductors transformer, cross-coupled capacitor. |
| [50] (2021) | 65 | CMOS_Bulk | 3 Pseudo-diff. CS | No | 95 | 9 | 20.5 | 13.4 | 11.6 | 9.3 | 1.2 | 170 | 0.025 | – | 0.8 V (Nominal) biasing, stacked inductors transformer, cross-coupled capacitor. |
| [29] (2022) | 65 | CMOS_Bulk | 4 Diff CS | 128-way | 98 | 9.7 | 32.5 | 32.1 | 15 | 28 | 1 | – | 4.35 | – | Center- and side-fed coupled-line (CL) based power combine scheme and neutralisation. |
| [51] (2022) | 28 | CMOS_FD_SOI | 2 Pseudo-diff. Triple-stacked-FET | 2-way | 99 | 10.3 | 21.8 | 15.1 | 18.6 | 12.3 | 3 | – | 0.054 | 89.5 | Triple-stacked-FET, custom transistor unit layout, cross-coupled capacitor |
| [52] (2016) | 65 | CMOS_Bulk | 4 Diff. CS | 2-way | 109 | 17 | 20.3 | 15.2 | 10.3 | 12.5 | 1.2 | – | 0.103/0.34* | 86.4 | Transformer-based current combiner, cross-coupled capacitor, diode linearizer |
| [53] (2022) | 16 | CMOS_FinFET | 2 Diff. | No | 110 | – | 17.3 | 10.5 | 22.6 | 7.3 | 0.85 | 46.7 | 0.023 | – | Skip layer vias, continuous via, two layers staked coil transformer, cross-coupled capacitor |
| [53] (2022) | 16 | CMOS_FinFET | 2 Diff. | No | 110 | – | 17.1 | 11.8 | 23.8 | 9.2 | 1 | 58.9 | 0.023 | – | Skip layer vias, continuous via, two layers staked coil transformer, cross-coupled capacitor |
| [54] (2018) | 65 | CMOS_Bulk | 4 Diff. CS | 2-way | 118 | 17 | 22.3 | 14.5 | 10.2 | 12.2 | 1.2 | – | 0.103/0.34* | 88.3 | Transformer-based current combiner, cross-coupled capacitor, diode linearizer |
| [55] (2019) | 40 | CMOS | 3 Diff. CS | 4-way | 120 | 38.5 | 16 | 14.6 | 9.4 | 9.3 | 1.2 | – | 0.1/0.33* | 500.7† | Pole-controlled transformer based IMN, transformer-based current combiner, capacitive neutralisation, cold-FET linearizer |
| [56] (2018) | 40 | CMOS | 3 Diff. CS | 2-way | 122 | 40 | 15.4 | 13.5 | 9 | 6.5 | – | – | – | – | On-chip transformer current combining, cross-coupled capacitor, cold-FET linearizer |
| [57] (2019) [58], (2020) | 22 | CMOS_Bulk | 4 Diff. CS | No | 132 | 22 | 22.5 | 8 | 6.6 | 5.2 | 0.9 | – | 0.0265 | 81.1 | Transformer-based matching network (TMN), cross-coupled capacitance |
| [59] (2019) | 65 | CMOS_Bulk | 3 Diff. Cascode | No | 133.2 | 22 | 10.5 | 9 | 5.2 | 4.2 | 2 | – | 0.275* | | Transformer cascaded, π-matching network, cross-coupled capacitance |

*(continued on next page)*

**Table 5** (*continued*)

| Ref. | Process (nm) | Technology | Topology | Power Combining | fc (GHz) | BW3dB (GHz) | Gain (dB) | Psat (dBm) | PAEpeak (%) | OP1dBm (dBm) | VDD (V) | Power (mW) | Core Area (mm2) | FoM** | Key Circuit Techniques |
|---|---|---|---|---|---|---|---|---|---|---|---|---|---|---|---|
| [61] (2022) | 45 | CMOS_SOI | 4 Diff. Cascode | 4-way | 135 | 15 | 24.8 | 18.5 | 11 | 13.5 | 2.4 | – | 0.46 | 96.3 | Stacking transistor core, inductor feedback, cross-coupled capacitance |
| [62] (2022) | 22 | CMOS_FD_SOI | 3 Diff. CS | 2-way | 135 | 22 | 13 | 8.5 | 7.4 | 7.6 | 0.8 | – | 0.157 | – | Dynamic bias-scaling, cascade, cross-coupled capacitance |
| [62] (2022) | 22 | CMOS_FD_SOI | 3 Diff. CS | 2-way | 135 | 20 | 14.2 | 10.3 | 7.9 | 9.6 | 1 | – | 0.157 | – | Dynamic bias-scaling, cascade, cross-coupled capacitance |
| [63] (2021) | 28 | CMOS_Bulk | 3 Pseudo-diff. CS | 2-way | 135 | 20 | 21.9 | 11.8 | 10.7 | 7.5 | 1 | 140 | 0.24/ 0.354* | 86.6 | Low-k transformer, artificial transmission lines, magnetic coupled-resonator matching network, cross-coupled capacitance |
| [64] (2020) | 16 | CMOS_Fin_FET | 3 Diff. CS | 2-way | 135 | 22 | 19 | 13.1 | 11 | 7.1 | 0.8 | 162 | 0.041/ 0.062* | 85.1 | Tileable small transistor, high-k sandwiched transformer, cross-coupled capacitance |
| [64] (2020) | 16 | CMOS_Fin_FET | 3 Diff. CS | 2-way | 135 | 22 | 20.5 | 15 | 12.8 | 9.2 | 1 | 207 | 0.041/ 0.062* | 89.2 | Tileable small transistor, high-k sandwiched transformer, cross-coupled capacitance |
| [65] (2022) | 28 | CMOS_Bulk | 3 Diff. | 4-way | 136 | 28 | 22.6 | 16.2 | 8.6 | 11.4 | 1 | – | 0.33/ 0.368* | 90.9 | Low-k transformer based wideband matching, magnetically resonator coupled network, Chebyshev artificial TL based power combiner, cross-coupled capacitance |
| [61] (2022) [60], (2021) | 45 | CMOS_SOI | 4 Diff. CS | 4-way | 140 | 21 | 22.2 | 16 | 12.5 | 12.5 | 1 | 305 | 0.43/ 0.46* | 92.2 | TL-combiner, immediate-k transformer, cross-coupled capacitance |
| [61] (2022) [60], (2021) | 45 | CMOS_SOI | 4 Diff. CS | 4-way | 140 | 21 | 24 | 17.5 | 13.4 | 14.2 | 1.2 | 410 | 0.43/ 0.46* | 95.7 | TL-combiner, immediate-k transformer, cross-coupled capacitance |
| [66] (2021) [67], (2022) | 22 | CMOS_FD_SOI | 4 Diff. CS | No | 140 | 30 | 33.6 | 12.5 | 10.8 | 9.4 | – | 152 | – | – | Double stacked-FET power stage, sandwiched-like interstage transformer matching network, cross-coupled capacitance |
| [68] (2020) | 65 | CMOS_Bulk | 1 Pseudo-diff. CS | 2-way | 140 | – | 5.7 | 11.4 | 2.5 | 6 | 2 | 100 | 0.082/ 0.395* | – | Asymmetric balun-based power combiner, cross-coupled capacitance |
| [68] (2020) | 65 | CMOS_Bulk | 2 Diff. CS | 4-way | 140 | – | 1 | 14 | 0.09 | 9.4 | 2 | 578 | 0.57/ 1.23* | – | Symmetric balun-based power combiner, cross-coupled capacitance |

**Table 5** (*continued*)

| Ref. | Process (nm) | Technology | Topology | Power Combining | fc (GHz) | BW3dB (GHz) | Gain (dB) | Psat (dBm) | PAEpeak (%) | OP1dBm (dBm) | VDD (V) | Power (mW) | Core Area (mm2) | FoM** | Key Circuit Techniques |
|---|---|---|---|---|---|---|---|---|---|---|---|---|---|---|---|
| [69] (2018) | 40 | CMOS_Bulk | 3 Pseudo-diff. Cascode | 2-way | 140 | 17 | 20.3 | 14.8 | 8.9 | 10.7 | 1 | 305 | 0.125/0.34* | 87.5 | Stacked structure transformer, optimized GSG pad, cross-coupled capacitance |
| [70] (2016) | 65 | CMOS | 3 Diff. CS | No | 140 | 25 | 16.8 | 7.9 | 7.5 | 2.5 | 1.2 | 80 | 0.2/0.564* | – | Cascode last stage, cross-coupled capacitance |
| [85] 2024 | 28 | CMOS_Bulk | 4 differential CS | No | 140 | 60 | 21.9 | 9.5 | 7.6 | 4.2 | 0.9 | – | 0.042 | – | Optimized CS stage, neutralising capacitor, transformer based matching network |
| [86] 2024 | 28 | CMOS_Bulk | 3 single-ended CS | 2-way | 140 | 16.51dB | 19.2 | 15.4 | 14.25 | 11.2 | 1 | 235 | 0.091 | 88.9 | De-overstabilization technique, neutralisation capacitor, distributed active transformer power combining |
| [71] (2021) | 65 | CMOS_Bulk | 2 Single-ended CS | No | 150 | 5 | 17.5 | 9.4 | 13.3 | 4 | 1 | 52.4 | 0.202 | – | Gmax based simultaneous gain and output power matching, Optimized three-passive-elements-based embedding network |
| [71] (2021) | 65 | CMOS_Bulk | 2 Single-ended CS | No | 150 | 5 | 17.5 | 10.4 | 16.1 | 5.3 | 1.2 | 86.3 | 0.202 | – | Gmax based simultaneous gain and output power matching, Optimized three-passive-elements-based embedding network |
| [72] (2021) | 45 | CMOS_SOI | 4 Pseudo-diff. CS | No | 152.5 | 17.5 | 18 | 8.8 | 6.8 | 5 | 1 | 92 | 0.04 | – | Flipped and orthogonal transformer with vertical coupling, cross-coupled capacitance |
| [72] (2021) | 45 | CMOS_SOI | 4 Pseudo-diff. CS | No | 167.5 | 14.‡ | 15.5 | 6.5 | 4.7 | 3 | 1 | 95 | 0.04/0.281* | – | Flipped and orthogonal transformer with vertical coupling, cross-coupled capacitance |
| [73] (2020) | 55 | CMOS | 5 Single-ended CS | No | 170 | 18 | 24 | 9 | 10 | 4.3 | 1.2 | 78.5 | 0.5 | – | Positive voltage feedback using coplanar waveguide (CPW) based inductors, coupled grounded-CPW, reverse coupled CPW lines, $\pi$-matching network |
| [74] (2019) | 40 | CMOS | 5 Pseudo-diff. CS | No | 195 | 32.8 | 9.6 | 3 | – | −4.3 | 0.9 | 51.8 | – | – | Coupled transmission line, staggered interstage wideband matching, cross-coupled capacitance |
| [75] (2020) | 65 | CMOS_Bulk | 8 Cascode | 2-way | 202 | 13.4 | 19.5 | 9.4 | 1.03 | 6.3 | 2.4 | 732 | 0.92 | – | Embedded matched-cascode amp-cell, slot power combiner |

**Table 5** (*continued*)

| Ref. | Process (nm) | Technology | Topology | Power Combining | fc (GHz) | BW3dB (GHz) | Gain (dB) | Psat (dBm) | PAEpeak (%) | OP1dBm (dBm) | VDD (V) | Power (mW) | Core Area (mm2) | FoM** | Key Circuit Techniques |
|---|---|---|---|---|---|---|---|---|---|---|---|---|---|---|---|
| [76] (2021) | 65 | CMOS_Bulk | 8 Pseudo-diff. CS | 4-way | 204 | 6 | 8.8 | 6.9 | 0.9 | – | – | 520 | 0.17/ 0.449* | – | Anti-phase stagging, cross-coupled capacitance |
| [77] (2021) | 65 | CMOS_Bulk | 6 Single-ended CS | 4-way | 243 | – | 26 | 9.3 | 2.8 | 6.5 | 1 | 300 | 1.038* | – | Gmax and TL based zero-degree splitters/ combiners, three passive-elements based embedding network |
| [77] (2021) | 65 | CMOS_Bulk | 6 Single-ended CS | 4-way | 243 | – | 26 | 10.5 | 2.7 | 7.7 | 1.2 | 407 | 1.038* | – | Gmax and TL based zero-degree splitters/ combiners, three passive-elements based embedding network |
| [77] (2021) | 65 | CMOS_Bulk | 6 Single-ended CS | 2-way | 245 | – | 28 | 8.6 | 4.6 | 5 | 1 | 166 | 0.488* | – | Gmax and TL based zero-degree splitters/ combiners, three passive-elements based embedding network |
| [77] (2021) | 65 | CMOS_Bulk | 6 Single-ended CS | 2-way | 245 | – | 28 | 9.2 | 3.9 | 6 | 1.2 | 217 | 0.488* | – | Gmax and TL based zero-degree splitters/ combiners, three passive-elements based embedding network |
| [78] (2021) | 65 | CMOS_Bulk | 3 Single-ended CS | No | 248.6 | – | 12.6 | 3 | 4.8 | 1.4 | 1 | 40 | 0.409* | – | Double- Gmax gain boosting, L-matching, three passive-elements based embedding network Gmax core |
| [78] (2021) | 65 | CMOS_Bulk | 3 Single-ended CS | No | 248.6 | – | 12.6 | 3.8 | 3.2 | 2.3 | 1.2 | 62.4 | 0.409* | – | Double- Gmax gain boosting, L-matching, three passive-elements based embedding network Gmax core |

* With RF-pad.
** FoM = Gain[dB]+Psat[dBm]+20*log(fc[GHz])+10*log(PAEmax[%]).
† FoM = Pout (W) x G (abs) x PAE (%) x Freq2(GHz) x Frac. BW (%).
⁂Hardware limitation.

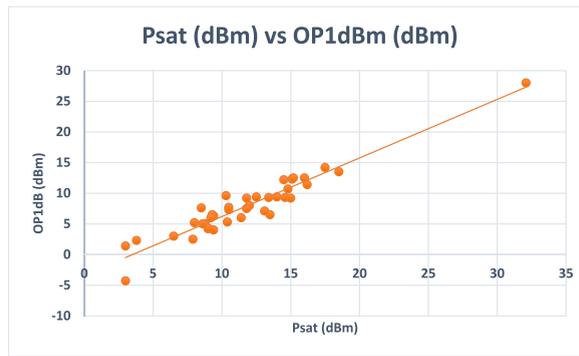

**Fig. 37.** PA performance trend of output power 1 dB compression point based on output saturated power.

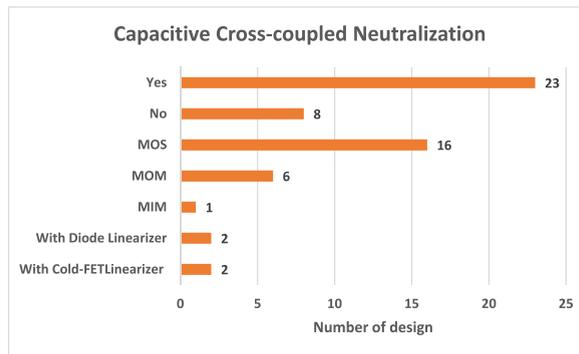

**Fig. 38.** Capacitive cross-coupled neutralisation interest for sub-THz band CMOS PA.

(4) articles implemented Linearisers along with cross-coupled neutralisations: two being Diode Linearisers and the remaining two being Cold-FET Linearisers. Amongst the 23 articles that utilised capacitive cross-coupled neutralisation techniques, 16 of them adopted MOS capacitors, six (6) implemented MOM capacitors and remaining one (1) used metal-insulator-metal (MIM) capacitor. All the differential PA designs adopt the capacitor neutralisation method. MOS transistor capacitor gained the highest interest due to its robust neutralisation for better matching with CMOS differential pair [83]. MOS capacitor may limit the PA linearity caused by capacitance variation between the voltages. Designs presented in Refs. [52,56] include a diode and Cold-FET pre-distortion lineariser to overcome the limitations [84]. MOM capacitors have the advantages of better linearity and compactness [50,55]. MIM capacitors are able to improve the PA quality factor [76]. However, some design does not utilise the capacitive cross-coupled neutralisation method due to the poor gain boosting effect caused by the lack of independent controlling of the phase and magnitude difference of the transistor gate and drain voltage [29]. Only capacitive cross-coupled neutralisation is easier to design for differential PA topologies, and it possesses a lack of gain boosting effect. Therefore, researchers use this method alongside other design specifications from the perspectives of lineariser, layout, pad and transformer, to resolve the parasitic effect and enhance the boosting effect.

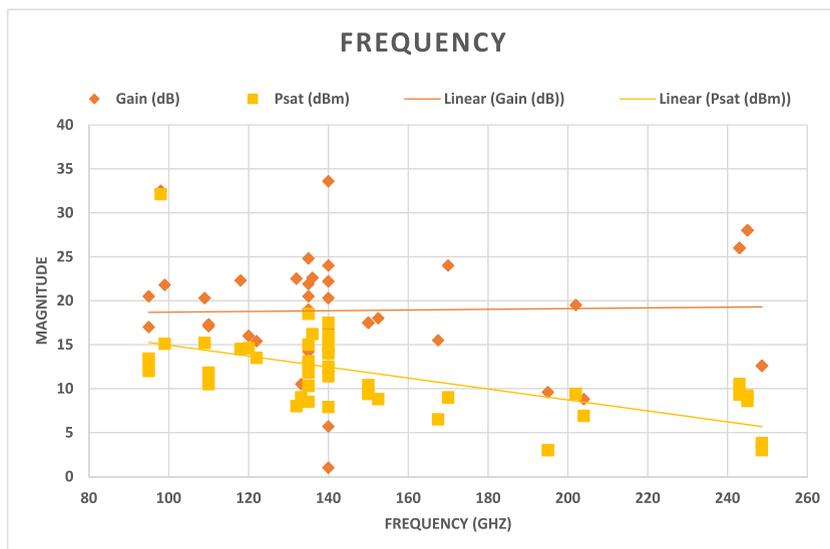

**Fig. 39.** PA linear gain and saturated output power, $P_{sat}$ trend over the frequency.

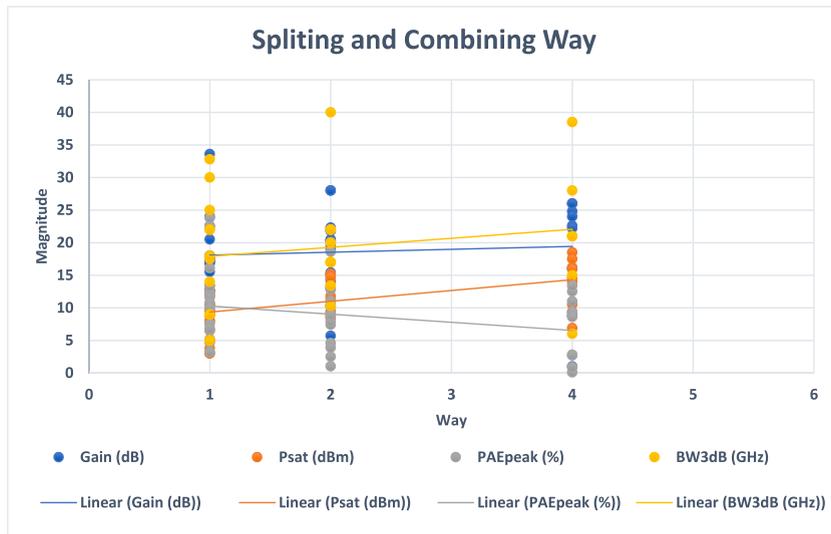

**Fig. 40.** PA performance trend of linear gain, saturated output power, bandwidth and efficiency over the power combining ways.

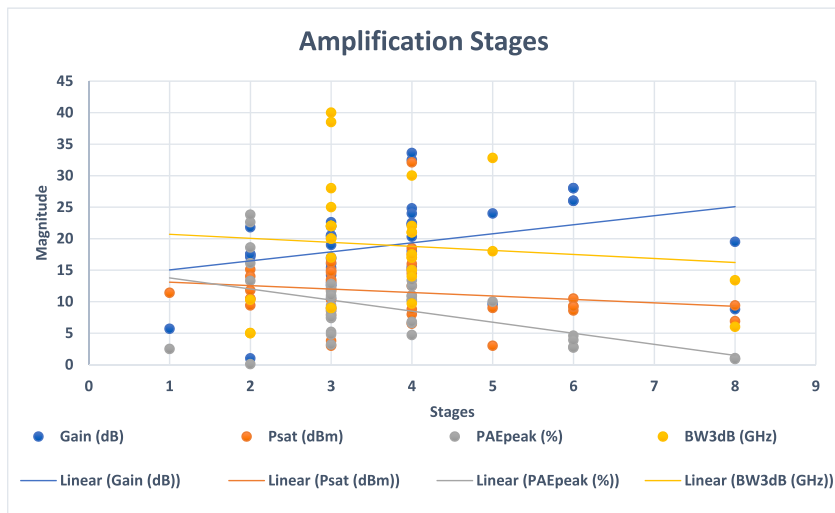

**Fig. 41.** PA performance trend of linear gain, saturated output power, bandwidth and efficiency over the amplification stages.

Figs. 39–41 summarised the sub-THz CMOS PA design trends based on the available reported designs, to serve as a reference. Fig. 39 shows that the saturated output power trend will deteriorate as the frequency increase due to Johnson's Limit, whereas the linear gain trend of the PA remains almost stable within the range of the plot. At higher frequencies, the gain of the PA is affected by mostly due to parasitic effect, miller effect and impedance matching issues; however, these effects are compensated by adopting additional circuitries, such as gain boosting module, customised inter-stage impedance matching, capacitance neutralisation and adjusting splitting-combining ways, etc. In addition, as there is a direct relationship between the saturated output power and the gain of a PA, the saturated power performance also deteriorates with frequency. It is possible to easily improve the saturated output power by simply raising the power or current splitting and combining ways, as shown in Fig. 40, which demonstrates the splitting/combining ways on gain, saturated power, pawer added efficiency and 3 dB bandwidth. In higher frequencies, PA is able to attain higher gain value, as shown in Fig. 39. The PA gain also shows an average linear gain trend of around 20 dB over the sub-THz band. The miller effect has been solved by increasing the number of stages of the PA, as shown in Fig. 41. While the gain trend increased with the

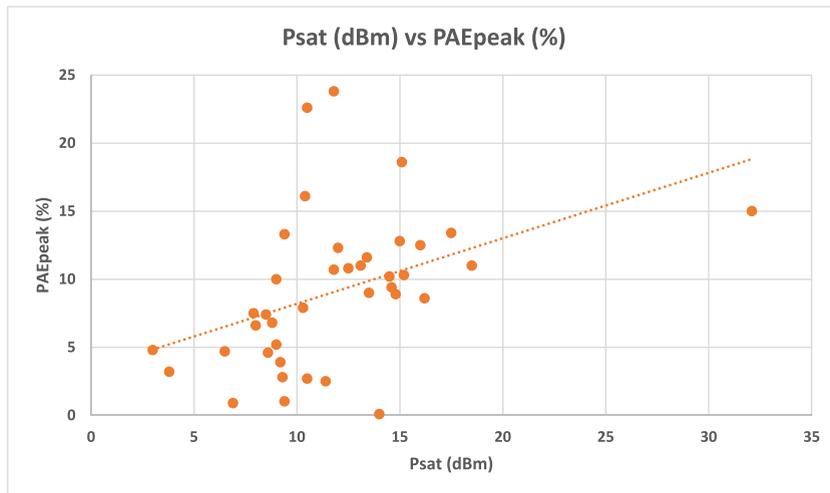

**Fig. 42.** PA performance trend of efficiency based on output saturated power, $P_{sat}$.

number of amplified stages because of miller effect the output saturated power, 3 dB bandwidth and power added efficiency decreased. Along with appropriate capacitive neutralisation technique and impedance matching circuitries, a minimum of three amplification stages are required to reach the satisfying overall performance trade-off, including power, driver, and/or neutralisation stages. Fig. 42 illustrates the relationship between power saturation (x-axis) and peak power added efficiency (y-axis). PAE inherently peaks around saturation; however, it declines if the PA is overdriven beyond its compression point. It has been observed that the maximum efficiency is achieved when the output power reached saturation; the trend remains positive as a high output power PA design is also able to achieve relatively higher efficiency. Therefore, careful selection of the stages and splitting/combining ways is needed to realise the respective design objectives in different applications for the 6G communications system.

## 6. Conclusion

In this review, we have investigated the state-of-the-art CMOS PA designs working in the sub-THz band. CMOS PA in sub-THz faces many parasitic and technological limitations. However, by adopting different design topologies, is it still possible to surpass the limitations and ensure stable performance. For future 6G communications, CMOS PA can be an important block in the transceiver to ensure signal emission stability while maintaining low cost, compared to other technologies. With the support of 6G communication, various cutting-edge applications can be realised, such as AR/VR applications, digital twins, unmanned aerial vehicles, autonomous driving and cloud computing which demand high transmission rate and low latency. D-band PA remains the most popular option for designing a sub-THz CMOS PA with a differential CS topology. Although the linear gain can maintain a 20 dBm trend over the sub-THz band, output saturated power still follows Johnson's limit trend and the PAE remains the greatest challenge for PA design. In short, researchers may focus on developing a high output power and efficient PA by integrating possible key circuit technologies to achieve the design objectives, such as power combining, multiple-stages, coupled transmission line and $G_{max}$-based with embedded networks based on three passive elements that show satisfying results. As a reference, this review will serve as a guideline to design a sub-THz CMOS PA and choose the appropriate topology for various design objectives and requirements based on future 6G wireless communication applications.

## CRediT authorship contribution statement

**Jun Yan Lee:** Conceptualization, Data curation, Methodology, Validation, Writing – original draft. **Duo Wu:** Methodology, Validation, Writing – original draft. **Xuanrui Guo:** Investigation, Validation, Visualization. **Jian Ding Tan:** Investigation, Methodology. **Teh Jia Yew:** Funding acquisition, Supervision. **Zi Neng Ng:** Investigation, Methodology. **Mohammad Arif Sobhan Bhuiyan:** Project administration, Supervision, Validation, Writing – review & editing, Funding acquisition. **Mahdi H. Miraz:** Supervision, Validation, Writing – review & editing, Funding acquisition.

## Data and code availability statement

No new data was generated for the research described in the article.

## Declaration of competing interest

The authors declare that they have no known competing financial interests or personal relationships that could have appeared to influence the work reported in this paper.

## Acknowledgment

The authors would like to acknowledge the support from Xiamen University Malaysia Research Fund under project codes XMUMRF/2020-C6/IECE/0016, XMUMRF/2021-C8/IECE/0021 and XMUMRF/2021-C8/IECE/0025.

## Appendix A. Supplementary data

Supplementary data to this article can be found online at https://doi.org/10.1016/j.heliyon.2025.e43393.

## References

[1] S. Chen, S. Kang, A tutorial on 5G and the progress in China, Frontiers of Information Technology & Electronic Engineering 19 (3) (2018) 309–321, https://doi.org/10.1631/FITEE.1800070.

[2] Y. Huang, Y. Shen, J. Wang, From terahertz imaging to terahertz wireless communications, Engineering (2022), https://doi.org/10.1016/j.eng.2022.06.023.

[3] F. Kausar, F.M. Senan, H.M. Asif, K. Raahemifar, 6G technology and taxonomy of attacks on blockchain technology, Alex. Eng. J. 61 (6) (2022) 4295–4306, https://doi.org/10.1016/j.aej.2021.09.051.

[4] D.C. Nguyen, et al., 6G internet of things: a comprehensive survey, IEEE Internet Things J. 9 (1) (2022) 359–383, https://doi.org/10.1109/JIOT.2021.3103320.

[5] U. Gustavsson, et al., Implementation challenges and opportunities in Beyond-5G and 6G communication, IEEE Journal of Microwave 1 (1) (2021) 86–100, https://doi.org/10.1109/JMW.2020.3034648.

[6] Z. Qadir, K.N. Le, N. Saeed, H.S. Munawar, Towards 6G internet of things: recent advances, use cases, and open challenges, ICT Express (2022), https://doi.org/10.1016/j.icte.2022.06.006.

[7] S.K. Pattnaik, et al., Future wireless communication technology towards 6G IoT: an application-based analysis of IoT in real-time location monitoring of employees inside underground mines by using BLE, Sensors 22 (9) (2022), https://doi.org/10.3390/s22093438.

[8] J.Y. Lee, D. Wu, X. Guo, M. Ariannejad, M.A.S. Bhuiyan, M.H. Miraz, Design of a W-band High-PAE class A & AB power amplifier in 150 nm GaAs technology, Transactions on Electrical and Electronic Materials 25 (3) (2024) 304–313, https://doi.org/10.1007/s42341-024-00513-8.

[9] M.W. Akhtar, S.A. Hassan, R. Ghaffar, H. Jung, S. Garg, M.S. Hossain, The shift to 6G communications: vision and requirements, Human-centric Computing and Information Sciences 10 (1) (2020) 53, https://doi.org/10.1186/s13673-020-00258-2.

[10] H. Tataria, M. Shafi, A.F. Molisch, M. Dohler, H. Sjöland, F. Tufvesson, 6G wireless systems: Vision, requirements, challenges, insights, and opportunities, in: Proceedings of the IEEE, 109, 2021, pp. 1166–1199, https://doi.org/10.1109/JPROC.2021.3061701, 7.

[11] T. Eichler, R. Ziegler, White paper: fundamentals of THz technology for 6G, Available: https://www.rohde-schwarz.com/premiumdownload/MGoxcjhJZTN0UzJmVjZ4VVZtQkdVTE5NTGZaZWNqbHR4OTB5THJEWnBZUGJob2g1NEU2NHVXT3RxN1pkVlQ4Y1ZuQXRVV3hub2hFVzRieTdSTlpkUVZiYStQbitFTkdLZUJMaXlNUC81Z1g1alpoeENkczhiSm5zM3R5RDlieVVMdmlXSXBDWUtrMjRHR0dtdG9rV1ZHU0thNmNYZCtOb2NzMlQ4U2lQRFk5Y1gzM3ZMSEtnTVFpRkxEMElEVVk2WFhsaUN1Sjh0bGRPOFhFdnQzelVYU3hNcHkrWllYaG0vTnc9OjphNDJiNjg3OTZlYlMjJNGMxM2U0NDZiODZkMDI1NTJiNw%3D%3D. (Accessed 13 October 2022).

[12] IEEE standard for high data rate wireless multi-media networks–amendment 2: 100 Gb/s wireless switched point-to-point physical layer, IEEE Std 802 (2017) 1–55, https://doi.org/10.1109/IEEESTD.2017.8066476, 15.3d-2017 (Amendment to *IEEE Std 802.15.3-2016 as amended by IEEE Std 802.15.3e-2017*.

[13] V. Petrov, T. Kurner, I. Hosako, IEEE 802.15.3d: first standardization efforts for sub-terahertz band communications toward 6G, IEEE Commun. Mag. 58 (11) (2020) 28–33, https://doi.org/10.1109/MCOM.001.2000273.

[14] N. Grace, FCC TAKES STEPS TO OPEN SPECTRUM HORIZONS FOR NEW SERVICES AND TECHNOLOGIES, vol. 15, FCC News, Washington, Mar, 2019 [Online]. Available: https://www.fcc.gov/document/fcc-opens-spectrum-horizons-new-services-technologies. (Accessed 13 October 2022).

[15] "White Paper on Network 2030 - A Blueprint of Technology, Applications and market drivers towards the year 2030 and beyond," [Online]. Available: https://www.itu.int/en/ITU-T/focusgroups/net2030/Documents/White_Paper.pdf, May 2019. (Accessed 13 October 2022).

[16] A. Pärssinen, et al., White Paper on RF Enabling 6G: Opportunities and Challenges from Technology to Spectrum, Oulu, Apr. 2021 [Online]. Available: http://urn.fi/urn:isbn:9789526228419. (Accessed 13 October 2022).

[17] 6G the next hyper-connected experience for all. https://cdn.codeground.org/nsr/downloads/researchareas/6G%20Vision.pdf, Jul. 2020. (Accessed 13 October 2022).

[18] 6G spectrum expanding the frontier. https://cdn.codeground.org/nsr/downloads/researchareas/2022May_6G_Spectrum.pdf, 2022. (Accessed 13 October 2022).

[19] "communications network. https://www-file.huawei.com/-/media/corp2020/pdf/giv/industry-reports/communications_network_2030_en.pdf, 2025. (Accessed 13 October 2022).

[20] H. Viswanathan, P.E. Mogensen, Communications in the 6G era, IEEE Access 8 (2020) 57063–57074, https://doi.org/10.1109/ACCESS.2020.2981745.

[21] China sends 'world's first 6G' test satellite into orbit, BBC News (Nov. 07, 2020). https://www.bbc.com/news/av/world-asia-china-54852131. (Accessed 12 October 2022).

[22] G.L. LaCombe, F. Dohi, Keysight Technologies First to Receive FCC Spectrum Horizons License for Developing 6G Technology in Sub-terahertz, Keysight, Santa Rosa, Mar. 08, 2022.

[23] H.-J. Song, N. Lee, Terahertz communications: challenges in the next decade, IEEE Trans Terahertz Sci Technol 12 (2) (2022) 105–117, https://doi.org/10.1109/TTHZ.2021.3128677.

[24] Z. Griffith, M. Urteaga, P. Rowell, A compact 140-GHz, 150-mW high-gain power amplifier MMIC in 250-nm InP HBT, IEEE Microw. Wireless Compon. Lett. 29 (4) (2019) 282–284, https://doi.org/10.1109/LMWC.2019.2902333.

[25] Y. Liu, et al., A G-band balanced power amplifier based on InP HEMT technology, in: 2020 IEEE MTT-S International Wireless Symposium (IWS), 2020, pp. 1–3, https://doi.org/10.1109/IWS49314.2020.9360136.

[26] M. Ćwikliński, et al., D-Band and G-Band high-performance GaN power amplifier MMICs, IEEE Trans. Microw. Theor. Tech. 67 (12) (2019) 5080–5089, https://doi.org/10.1109/TMTT.2019.2936558.

[27] F. Thome, A. Leuther, A 75–305-GHz power amplifier MMIC with 10–14.9-dBm pout in a 35-nm InGaAs mHEMT technology, IEEE Microw. Wireless Compon. Lett. 31 (6) (2021) 741–743, https://doi.org/10.1109/LMWC.2021.3058101.

[28] V.D. Tran, S. Chakraborty, J. Mihaljevic, S. Mahon, M. Heimlich, A W-band driver amplifier in 0.1 μm pHEMT gallium arsenide process, in: 2021 IEEE Asia-Pacific Microwave Conference (APMC), 2021, pp. 46–48, https://doi.org/10.1109/APMC52720.2021.9661675.

[29] W. Zhu, et al., A 1V 32.1 dBm 92-to-102GHz power amplifier with a scalable 128-to-1 power combiner achieving 15% peak PAE in a 65nm bulk CMOS process, in: 2022 IEEE International Solid- State Circuits Conference (ISSCC) vol. 65, 2022, pp. 318–320, https://doi.org/10.1109/ISSCC42614.2022.9731700.

[30] I. Petricli, D. Riccardi, A. Mazzanti, D-Band SiGe BiCMOS power amplifier with 16.8dBm P₁dB and 17.1% PAE enhanced by current-clamping in multiple common-base stages, IEEE Microw. Wireless Compon. Lett. 31 (3) (2021) 288–291, https://doi.org/10.1109/LMWC.2021.3049458.

[31] Y. Liu, Advantages of CMOS technology in very large scale integrated circuits, in: 2021 2nd International Conference on Artificial Intelligence in Electronics Engineering, 2021, pp. 82–88, https://doi.org/10.1145/3460268.3460280.

[32] Y. Taur, et al., CMOS scaling into the nanometer regime, in: Proceedings of the IEEE, vol. 85, 1997, pp. 486–504, https://doi.org/10.1109/5.573737, 4.

[33] K. Seshan, Chapter 2 - limits and hurdles to continued CMOS scaling, in: K. Seshan, D. Schepis (Eds.), Handbook of Thin Film Deposition, fourth ed., William Andrew Publishing, 2018, pp. 19–41, https://doi.org/10.1016/B978-0-12-812311-9.00002-5.

[34] G. Kumar, S. Agrawal, CMOS limitations and futuristic carbon allotropes, in: 2017 8th IEEE Annual Information Technology, Electronics and Mobile Communication Conference, IEMCON), 2017, pp. 68–71, https://doi.org/10.1109/IEMCON.2017.8117151.

[35] V.A. Bespalov, N.A. Dyuzhev, V. Yu Kireev, Possibilities and limitations of CMOS technology for the production of various microelectronic systems and devices, Nanobiotechnology Reports 17 (1) (2022) 24–38, https://doi.org/10.1134/S2635167622010037.

[36] M.H. Wakayama, Nanometer CMOS from a mixed-signal/RF perspective, in: 2013 IEEE International Electron Devices Meeting, 2013, pp. 17.4.1–17.4.4, https://doi.org/10.1109/IEDM.2013.6724648.

[37] A.A. Farid, A. Simsek, A.S.H. Ahmed, M.J.W. Rodwell, A broadband direct conversion transmitter/receiver at D-band using CMOS 22nm FDSOI, in: 2019 IEEE Radio Frequency Integrated Circuits Symposium (RFIC), 2019, pp. 135–138, https://doi.org/10.1109/RFIC.2019.8701730.

[38] K. Statnikov, J. Grzyb, B. Heinemann, U.R. Pfeiffer, 160-GHz to 1-THz multi-color active imaging with a lens-coupled SiGe HBT chip-set, IEEE Trans. Microw. Theor. Tech. 63 (2) (2015) 520–532, https://doi.org/10.1109/TMTT.2014.2385777.

[39] P. Hillger, J. Grzyb, S. Malz, B. Heinemann, U. Pfeiffer, A lens-integrated 430 GHz SiGe HBT source with up to −6.3 dBm radiated power, in: 2017 IEEE Radio Frequency Integrated Circuits Symposium (RFIC), 2017, pp. 160–163, https://doi.org/10.1109/RFIC.2017.7969042.

[40] R. Jain, P. Hillger, E. Ashna, J. Grzyb, U.R. Pfeiffer, A 64-Pixel 0.42-THz source SoC with spatial modulation diversity for computational imaging, IEEE J. Solid State Circ. 55 (12) (2020) 3281–3293, https://doi.org/10.1109/JSSC.2020.3018819.

[41] S. Daneshgar, J.F. Buckwalter, Compact series power combining using Subquarter-Wavelength baluns in silicon germanium at 120 GHz, IEEE Trans. Microw. Theor. Tech. 66 (11) (2018) 4844–4859, https://doi.org/10.1109/TMTT.2018.2867467.

[42] E. Camargo, J. Schellenberg, L. Bui, N. Estella, F-Band, GaN power amplifiers, in: 2018 IEEE/MTT-S International Microwave Symposium - IMS, 2018, pp. 753–756, https://doi.org/10.1109/MWSYM.2018.8439280.

[43] D. Simic, P. Reynaert, A 14.8 dBm 20.3 dB power amplifier for D-band applications in 40 nm CMOS, in: 2018 IEEE Radio Frequency Integrated Circuits Symposium (RFIC), 2018, pp. 232–235, https://doi.org/10.1109/RFIC.2018.8428981.

[44] P. Rodríguez-Vázquez, J. Grzyb, B. Heinemann, U.R. Pfeiffer, A 16-QAM 100-Gb/s 1-M wireless link with an EVM of 17% at 230 GHz in an SiGe technology, IEEE Microw. Wireless Compon. Lett. 29 (4) (2019) 297–299, https://doi.org/10.1109/LMWC.2019.2899487.

[45] P. Rodriguez-Vazquez, J. Grzyb, B. Heinemann, U.R. Pfeiffer, A QPSK 110-Gb/s polarization-diversity MIMO wireless link with a 220–255 GHz tunable LO in a SiGe HBT technology, IEEE Trans. Microw. Theor. Tech. 68 (9) (2020) 3834–3851, https://doi.org/10.1109/TMTT.2020.2986196.

[46] IEEE standard letter designations for radar-frequency bands, in: IEEE Std 521-2019 (Revision of IEEE Std 521-2002), 2020, pp. 1–15, https://doi.org/10.1109/IEEESTD.2020.8999849.

[47] K. Rawat, P. Roblin, S.K. Koul, Introduction to RF power amplifier design and architecture, in: K. Rawat, P. Roblin, S.K. Koul (Eds.), Bandwidth and Efficiency Enhancement in Radio Frequency Power Amplifiers for Wireless Transmitters, Springer International Publishing, Cham, 2020, pp. 1–106, https://doi.org/10.1007/978-3-030-38866-9_1.

[48] P.M. Asbeck, N. Rostomyan, M. Özen, B. Rabet, J.A. Jayamon, Power amplifiers for mm-Wave 5G applications: technology comparisons and CMOS-SOI demonstration circuits, IEEE Trans. Microw. Theor. Tech. 67 (7) (2019) 3099–3109, https://doi.org/10.1109/TMTT.2019.2896047.

[49] M.K. Kazimierczuk, Introduction, in: RF Power Amplifiers, John Wiley & Sons, Ltd, 2014, pp. 1–64, https://doi.org/10.1002/9781118844373.ch1.

[50] T. Elazar, E. Socher, 95GHz 13dBm IQ-combined PA in 65nm CMOS, in: 2020 50th European Microwave Conference, EuMC), 2021, pp. 182–185, https://doi.org/10.23919/EuMC48046.2021.9338067.

[51] K. Kim, K. Lee, S.-U. Choi, J. Kim, C.-G. Choi, H.-J. Song, A 97–107 GHz Triple-Stacked-FET power amplifier with 23.7dB peak gain, 15.1dBm PSAT, and 18.6% PAEMAX in 28-nm FD-SOI CMOS, in: 2022 IEEE Radio Frequency Integrated Circuits Symposium (RFIC), 2022, pp. 183–186, https://doi.org/10.1109/RFIC54546.2022.9863175.

[52] H.S. Son, J.Y. Jang, D.M. Kang, H.J. Lee, C.S. Park, A 109 GHz CMOS power amplifier with 15.2 dBm psat and 20.3 dB gain in 65-nm CMOS technology, IEEE Microw. Wireless Compon. Lett. 26 (7) (2016) 510–512, https://doi.org/10.1109/LMWC.2016.2574834.

[53] Q. Yu, J. Garrett, S. Hwangbo, G. Dogiamis, S. Rami, An F-Band power amplifier with skip-layer via achieving 23.8% PAE in FinFET technology, in: 2022 IEEE Radio Frequency Integrated Circuits Symposium (RFIC), 2022, pp. 179–182, https://doi.org/10.1109/RFIC54546.2022.9863118.

[54] H.S. Son, et al., A D-band CMOS power amplifier for wireless chip-to-chip communications with 22.3 dB gain and 12.2 dBm P1dB in 65-nm CMOS technology, in: 2018 IEEE Topical Conference on RF/Microwave Power Amplifiers for Radio and Wireless Applications (PAWR), 2018, pp. 35–38, https://doi.org/10.1109/PAWR.2018.8310061.

[55] H.S. Son, T.H. Jang, S.H. Kim, K.P. Jung, J.H. Kim, C.S. Park, Pole-controlled wideband 120 GHz CMOS power amplifier for wireless chip-to-chip communication in 40-nm CMOS process, IEEE Transactions on Circuits and Systems II: Express Briefs 66 (8) (2019) 1351–1355, https://doi.org/10.1109/TCSII.2018.2880308.

[56] C.J. Lee, et al., A 120 GHz I/Q transmitter Front-end in a 40 nm CMOS for wireless chip to chip communication, in: 2018 IEEE Radio Frequency Integrated Circuits Symposium (RFIC), 2018, pp. 192–195, https://doi.org/10.1109/RFIC.2018.8429019.

[57] X. Tang, A. Medra, J. Nguyen, K. Khalaf, B. Debaillie, P. Wambacq, Design of A D-band transformer-based neutralized Class-AB power amplifier in silicon technologies, in: 2019 17th IEEE International New Circuits and Systems Conference (NEWCAS), 2019, pp. 1–4, https://doi.org/10.1109/NEWCAS44328.2019.8961277.

[58] X. Tang, et al., Design of D-Band transformer-based gain-boosting Class-AB power amplifiers in silicon technologies, IEEE Transactions on Circuits and Systems I: Regular Papers 67 (5) (2020) 1447–1458, https://doi.org/10.1109/TCSI.2020.2974197.

[59] G. Su, C. Wan, D. Chen, X. Gao, L. Sun, A 129.5-151.5GHz fully differential power amplifier in 65nm CMOS, in: 2019 IEEE MTT-S International Wireless Symposium (IWS), 2019, pp. 1–3, https://doi.org/10.1109/IEEE-IWS.2019.8803888.

[60] S. Li, G.M. Rebeiz, A 130-151 GHz 8-Way power amplifier with 16.8-17.5 dBm psat and 11.7-13.4% PAE using CMOS 45nm RFSOI, in: 2021 IEEE Radio Frequency Integrated Circuits Symposium (RFIC), 2021, pp. 115–118, https://doi.org/10.1109/RFIC51843.2021.9490507.

[61] S. Li, G.M. Rebeiz, High efficiency D-Band multiway power combined amplifiers with 17.5–19-dBm psat and 14.2–12.1% peak PAE in 45-nm CMOS RFSOI, IEEE J. Solid State Circ. 57 (5) (2022) 1332–1343, https://doi.org/10.1109/JSSC.2022.3145394.

[62] E. Rahimi, F. Bozorgi, G. Hueber, A 22nm FD-SOI CMOS 2-way D-band power amplifier achieving PAE of 7.7% at 9.6dBm OP1dB and 3.1% at 6dB back-off by leveraging adaptive back-gate bias technique, in: 2022 IEEE Radio Frequency Integrated Circuits Symposium (RFIC), 2022, pp. 175–178, https://doi.org/10.1109/RFIC54546.2022.9863164.

[63] J. Zhang, T. Wu, L. Nie, S. Ma, Y. Chen, J. Ren, A 120–150 GHz power amplifier in 28-nm CMOS achieving 21.9-dB gain and 11.8-dBm psat for Sub-THz imaging system, IEEE Access 9 (2021) 74752–74762, https://doi.org/10.1109/ACCESS.2021.3080710.

[64] B. Philippe, P. Reynaert, 24.7 A 15dBm 12.8%-PAE compact D-Band power amplifier with two-way power combining in 16nm FinFET CMOS, in: 2020 IEEE International Solid- State Circuits Conference - (ISSCC), 2020, pp. 374–376, https://doi.org/10.1109/ISSCC19947.2020.9062920.

[65] J. Zhang, T. Wu, Y. Chen, J. Ren, S. Ma, A 124–152 GHz > 15-dBm P_sat 28-nm CMOS PA using Chebyshev Artificial- transmission-line-based matching for wideband power splitting and combining, in: 2022 IEEE Radio Frequency Integrated Circuits Symposium (RFIC), 2022, pp. 187–190, https://doi.org/10.1109/RFIC54546.2022.9863134.

[66] X. Tang, J. Nguyen, G. Mangraviti, Z. Zong, P. Wambacq, A 140 GHz T/R front-end module in 22 nm FD-SOI CMOS, in: 2021 IEEE Radio Frequency Integrated Circuits Symposium (RFIC), 2021, pp. 35–38, https://doi.org/10.1109/RFIC51843.2021.9490470.

[67] X. Tang, J. Nguyen, G. Mangraviti, Z. Zong, P. Wambacq, Design and analysis of a 140-GHz T/R front-end module in 22-nm FD-SOI CMOS, IEEE J. Solid State Circ. 57 (5) (2022) 1300–1313, https://doi.org/10.1109/JSSC.2021.3139359.

[68] D. Yamazaki, T. Horikawa, T. Iizuka, A 140-GHz 14-dBm power amplifier using power combiner based on symmetric balun in 65-nm bulk CMOS, in: 2020 IEEE International Symposium on Radio-Frequency Integration Technology (RFIT), 2020, pp. 79–81, https://doi.org/10.1109/RFIT49453.2020.9226186.

[69] D. Simic, P. Reynaert, A 14.8 dBm 20.3 dB power amplifier for D-band applications in 40 nm CMOS, in: 2018 IEEE Radio Frequency Integrated Circuits Symposium (RFIC), 2018, pp. 232–235, https://doi.org/10.1109/RFIC.2018.8428981.

[70] X. Gao, L. Sun, J. Wen, G. Su, M. Jin, J. Zhou, A three stage, fully differential D-band CMOS power amplifier, in: 2016 13th IEEE International Conference on Solid-State and Integrated Circuit Technology, ICSICT, 2016, pp. 1558–1560, https://doi.org/10.1109/ICSICT.2016.7998801.

[71] D.-W. Park, D.R. Utomo, B. Yun, H.U. Mahmood, S.-G. Lee, A D-Band power amplifier in 65-nm CMOS by adopting simultaneous output power-and gain-matched gmax-core, IEEE Access 9 (2021) 99039–99049, https://doi.org/10.1109/ACCESS.2021.3096423.

[72] A. Hamani, A. Siligaris, B. Blampey, J.L.G. Jimenez, 167-GHz and 155-GHz high gain D-band power amplifiers in CMOS SOI 45-nm technology, in: 2020 15th European Microwave Integrated Circuits Conference (Eumic), 2021, pp. 261–264.

[73] X. Zhang, F. Meng, H. Fu, K. Ma, J. Ma, A G-band high-gain power amplifier with positive voltage-feedback in 55-nm CMOS technology, in: 2020 13th UK-Europe-China Workshop on Millimetre-Waves and Terahertz Technologies (UCMMT), 2020, pp. 1–3, https://doi.org/10.1109/UCMMT49983.2020.9296080.

[74] Y. Liu, T. Ma, P. Guan, L. Mao, B. Chi, A G-Band wideband bidirectional transceiver front-end in 40-nm CMOS, IEEE Transactions on Circuits and Systems II: Express Briefs 66 (5) (2019) 798–802, https://doi.org/10.1109/TCSII.2019.2908382.

[75] H. Bameri, O. Momeni, An embedded 200 GHz power amplifier with 9.4 dBm saturated power and 19.5 dB gain in 65 nm CMOS, in: 2020 IEEE Radio Frequency Integrated Circuits Symposium (RFIC), 2020, pp. 191–194, https://doi.org/10.1109/RFIC49505.2020.9218441.

[76] K.B. Atar, E. Socher, A 204 GHz power amplifier with 6.9dBm psat and 8.8dB gain in 65nm CMOS technology, in: 2021 IEEE International Conference on Microwaves, Antennas, Communications and Electronic Systems (COMCAS), 2021, pp. 173–177, https://doi.org/10.1109/COMCAS52219.2021.9629088.

[77] B. Yun, D.-W. Park, K.-S. Choi, H.-J. Song, S.-G. Lee, 245/243GHz, 9.2/10.5dBm saturated output power, 4.6/2.8% PAE, and 28/26dB gain power amplifiers in 65nm CMOS adopting 2-and 4-way power combining, in: 2021 IEEE Asian Solid-State Circuits Conference, A-SSCC), 2021, pp. 1–3, https://doi.org/10.1109/A-SSCC53895.2021.9634180.

[78] B. Yun, D.-W. Park, W.-J. Choi, H.U. Mahmood, S.-G. Lee, A 250-GHz 12.6-dB gain and 3.8-dBm psat power amplifier in 65-nm CMOS adopting dual-shunt elements based gmax-core, IEEE Microw. Wireless Compon. Lett. 31 (3) (2021) 292–295, https://doi.org/10.1109/LMWC.2020.3046745.

[79] D. Lee, A. Davydov, B. Mondal, G. Xiong, G. Morozov, J. Kim, From sub-terahertz to terahertz: challenges and design considerations, in: 2020 IEEE Wireless Communications and Networking Conference Workshops (WCNCW), 2020, pp. 1–8, https://doi.org/10.1109/WCNCW48565.2020.9124764.

[80] Y. Yu, J. Cai, X.-W. Zhu, P. Chen, C. Yu, Self-sensing digital predistortion of RF power amplifiers for 6G intelligent radio, IEEE Microw. Wireless Compon. Lett. 32 (5) (2022) 475–478, https://doi.org/10.1109/LMWC.2021.3139018.

[81] Q. Zhang, Y. Liu, J. Zhou, W. Chen, A complexity-reduced band-limited memory polynomial behavioral model for wideband power amplifier, in: 2015 IEEE International Wireless Symposium (IWS 2015), 2015, pp. 1–4, https://doi.org/10.1109/IEEE-IWS.2015.7164551.

[82] X. Li, et al., A 110-to-130GHz SiGe BiCMOS doherty power amplifier with Slotline-Based power-combining technique achieving >22dBm saturated output power and >10% power back-off efficiency, in: 2022 IEEE International Solid- State Circuits Conference (ISSCC), vol. 65, 2022, pp. 316–318, https://doi.org/10.1109/ISSCC42614.2022.9731552.

[83] M. Abdulaziz, H. v Hünerli, K. Buisman, C. Fager, Improvement of AM–PM in a 33-GHz CMOS SOI power amplifier using pMOS neutralization, IEEE Microw. Wireless Compon. Lett. 29 (12) (2019) 798–801, https://doi.org/10.1109/LMWC.2019.2948763.

[84] K.-Y. Kao, Y.-C. Hsu, K.-W. Chen, K.-Y. Lin, Phase-delay Cold-FET pre-distortion linearizer for millimeter-wave CMOS power amplifiers, IEEE Trans. Microw. Theor. Tech. 61 (12) (2013) 4505–4519, https://doi.org/10.1109/TMTT.2013.2288085.

[85] Z. Yan, et al., A D-Band power amplifier with 60-GHz large-signal bandwidth and 7.6% peak PAE in 28-nm CMOS, IEEE Microwave and Wireless Technology Letters 34 (5) (2024) 540–543, https://doi.org/10.1109/LMWT.2024.3380441.

[86] L. Zhang, et al., A compact 140-GHz power amplifier with 15.4-dBm psat and 14.25% peaking PAE in 28-nm bulk CMOS process, IEEE Trans. Microw. Theor. Tech. 72 (5) (2024) 3016–3030, https://doi.org/10.1109/TMTT.2023.3322742.